\documentclass[sigconf, review=false]{acmart}

\IfFileExists{upquote.sty}{\usepackage{upquote}}{}
\IfFileExists{microtype.sty}{
  \usepackage[]{microtype}
  \UseMicrotypeSet[protrusion]{basicmath} 
}{}
\makeatletter
\@ifundefined{KOMAClassName}{
  \IfFileExists{parskip.sty}{%
    \usepackage{parskip}
  }{
    \setlength{\parindent}{0pt}
    \setlength{\parskip}{6pt plus 2pt minus 1pt}}
}{
  \KOMAoptions{parskip=half}}
\makeatother

\PassOptionsToPackage{unicode}{hyperref}
\PassOptionsToPackage{hyphens}{url}
\PassOptionsToPackage{dvipsnames,svgnames,x11names}{xcolor}

\IfFileExists{bookmark.sty}{\usepackage{bookmark}}{\usepackage{hyperref}}

\usepackage{longtable,booktabs,array}
\usepackage{calc} 
\usepackage{etoolbox}
\makeatletter
\patchcmd\longtable{\par}{\if@noskipsec\mbox{}\fi\par}{}{}
\makeatother
\IfFileExists{footnotehyper.sty}{\usepackage{footnotehyper}}{\usepackage{footnote}}
\makesavenoteenv{longtable}
\usepackage{graphicx}
\makeatletter
\newsavebox\pandoc@box
\newcommand*\pandocbounded[1]{
  \sbox\pandoc@box{#1}%
  \Gscale@div\@tempa{\textheight}{\dimexpr\ht\pandoc@box+\dp\pandoc@box\relax}%
  \Gscale@div\@tempb{\linewidth}{\wd\pandoc@box}%
  \ifdim\@tempb\p@<\@tempa\p@\let\@tempa\@tempb\fi
  \ifdim\@tempa\p@<\p@\scalebox{\@tempa}{\usebox\pandoc@box}%
  \else\usebox{\pandoc@box}%
  \fi%
}
\def\fps@figure{htbp}
\makeatother

\usepackage[]{natbib}
\usepackage{xurl}
\subtitle{Privacy in Encryption Regulation and Lawful-Surveillance Protocols}
\setcctype{by-nc-nd}
\acmBooktitle{Symposium on Computer Science and Law (CSLAW '26), March 03--05, 2026, Berkeley, CA, USA}
\acmPrice{}
\definecolor{mypink}{RGB}{219, 48, 122}
\makeatletter
\@ifpackageloaded{caption}{}{\usepackage{caption}}
\AtBeginDocument{%
\ifdefined\contentsname
  \renewcommand*\contentsname{Table of contents}
\else
  \newcommand\contentsname{Table of contents}
\fi
\ifdefined\listfigurename
  \renewcommand*\listfigurename{List of Figures}
\else
  \newcommand\listfigurename{List of Figures}
\fi
\ifdefined\listtablename
  \renewcommand*\listtablename{List of Tables}
\else
  \newcommand\listtablename{List of Tables}
\fi
\ifdefined\figurename
  \renewcommand*\figurename{Figure}
\else
  \newcommand\figurename{Figure}
\fi
\ifdefined\tablename
  \renewcommand*\tablename{Table}
\else
  \newcommand\tablename{Table}
\fi
}
\@ifpackageloaded{float}{}{\usepackage{float}}
\floatstyle{ruled}
\@ifundefined{c@chapter}{\newfloat{codelisting}{h}{lop}}{\newfloat{codelisting}{h}{lop}[chapter]}
\floatname{codelisting}{Listing}

\makeatother
\makeatletter
\@ifpackageloaded{caption}{}{\usepackage{caption}}
\@ifpackageloaded{subcaption}{}{\usepackage{subcaption}}
\makeatother

\hypersetup{
  pdftitle={Security Is Not Enough},
  pdfauthor={Artur Pericles Lima Monteiro},
  colorlinks=true,
  linkcolor={blue},
  filecolor={Maroon},
  citecolor={Blue},
  urlcolor={red},
  pdfcreator={LaTeX via pandoc, via quarto}}

\AtBeginDocument{%
  }

\setcopyright{cc}
\copyrightyear{2026}
\acmYear{2026}
\acmDOI{10.1145/3788646.3789535}

\acmConference[CSLAW '26]{Symposium on Computer Science and Law}{March
3--5, 2026}{Berkeley, CA}
\acmPrice{}
\acmISBN{979-8-4007-2447-3/2026/03}

\begin{document}

\title{Security Is Not Enough}


  \author{Artur Pericles Lima Monteiro}
  \email{artur.monteiro@yale.edu}
  \orcid{0000-0002-8553-1219}
            \affiliation{%
                  \institution{Yale Jackson School of Global Affairs \&
Yale Law School}
                                  \city{New Haven}
                                  \country{USA}
                      }

\begin{abstract}
This article argues that security is not enough to fully capture what is
at stake in government exceptional access to encrypted data. A
conception of privacy as security has little to say about
``lawful-surveillance protocols''---an active research agenda in
cryptography that aims to enable government exceptional access without
compromising systemic security. But the limitations are not contingent
on the success of this agenda. The normative landscape today cannot be
explained if security is all there is to privacy. And fundamental
objections to Apple's abandoned client-side scanning system gesture
beyond security. This article's contribution is modest: to show that
there must be more to privacy than the security mold it has taken. A
richer understanding is needed both to assess policy and to guide
research on lawful-surveillance protocols.    
\end{abstract}



\maketitle

\setlength{\parskip}{-0.1pt}

\section*{Introduction}\label{introduction}
\addcontentsline{toc}{section}{Introduction}

Encryption protects privacy, but it can keep law enforcement from
information it legitimately needs. Should exceptional access mechanisms
be created so that officials can obtain encrypted data when they are
authorized by law? The opposition to exceptional access has converged on
a reframing of the dispute. It shifts the terms from security
vs.~privacy to security vs.~security. Under this view, exceptional
access should be rejected because it would undermine the very mission of
law enforcement, as any such mechanisms would also unreasonably
compromise citizens' and national security. A burgeoning research
agenda---lawful-surveillance protocols---aims to use cryptography to
enable exceptional access without compromising security.

This article argues that security is not enough to fully capture what is
at stake in government exceptional access to encrypted data. It has
little to say about lawful-surveillance protocols. Indeed, we see this
lurking in the response to Apple's proposed and abandoned client-side
scanning system by long-time critics of exceptional access. Yet
privacy-as-security's limitations are not just prospective and
contingent on the success of that research agenda. I argue that the
current normative landscape cannot be explained if security is all there
is to privacy. My contribution is modest: I aim only to show that there
must be more to privacy than the security argument whose mold it has
taken in encryption policy. I do not articulate here what an alternative
conception of the right to privacy would be.

Section~\ref{sec-background} briefly recapitulates the ``Crypto Wars.''
Section~\ref{sec-securing} summarizes and contextualizes the
privacy-as-security argument. Section~\ref{sec-surveillance-protocols}
surveys and assesses lawful-surveillance protocol proposals.
Section~\ref{sec-beyond-security} shows that not only is
privacy-as-security insufficient to evaluate such protocols, but also
that it does not explain why exceptional access mechanisms cannot offset
any introduced risks by modifying background institutional conditions,
such as statutory requirements for interception
(Section~\ref{sec-tradingsecurity}). It also fails to justify the
exceptional-access positions opponents have taken on government hacking
(Section~\ref{sec-govthacking}). In fact, the objections by opponents to
Apple's proposed client-side scanning system themselves require an
understanding of privacy that goes beyond security
(Section~\ref{sec-apple-csam}).

\section{The Long Crypto Wars}\label{sec-background}

Disputes about the legal status of cryptography, what it is and what it
should be, are long-standing. Tensions trace back at least to the very
first days of public-key cryptography \citep{Diffie:1976dc} in the 1970s
\citep{Vidan.2025cg}. A letter from an NSA official to the IEEE warned
that an October 1977 IEEE symposium on information theory would violate
the International Traffic in Arms Regulations \citep{Shapley.1977cs}.
The NSA stated that the official had acted in his personal capacity
\citep{USSen.1978de}, and the symposium went ahead.\footnote{Though,
  acting on advice from Stanford counsel, Martin Hellman presented the
  paper, instead of his then students Steve Pohlig and Ralph Merkle, as
  had been planned, as ``Hellman had the benefit of the tenure system to
  support him if he were to get into any trouble''
  \citep[p.~368]{West.2022ci}.} The NSF temporarily suspended new
funding for cryptographic research after pressure from the NSA
\citep[p.~351]{Bamford.1983pp}. Even Kahn's 1967 public-facing
\emph{Codebreakers} \citep{Kahn:1967tk} was subject to deletion requests
from the NSA, and before the Defense Department had insisted publication
``would not be in national interest''
\citep[p.~23]{Levy:2002wp}.\footnote{Levy notes the suppressed passages
  struck Kahn as unimportant \citep[p.~23]{Levy:2002wp}. Recent research
  \citep{Sherman.2023cw} shows the objections to the passages were
  raised by British intelligence, with whom the NSA shared the
  manuscript.}

Though the tensions did not start then, the term ``Crypto Wars''
referred to the contestation that arose in the 1990s, when encryption
regulation made headlines (Section~\ref{sec-clipper}). At that time,
national security and law enforcement officials worried that
commercially available encryption would soon take hold of communications
and seriously obstruct their mission. Since then, as strong encryption
was rolled out and turned on by default to billions of users, efforts to
restrict encryption have been championed by law enforcement, with
intelligence officials sometimes explicitly
\citep{Ashbrook.2016mh, Rogers.2016ce} siding with encryption as a
matter of national security (Section~\ref{sec-applevfbi}). More
recently, calls have also been made by other actors and for reasons not
limited to law enforcement or national security
(Section~\ref{sec-cryptowarsiii}).

This section offers a brief recapitulation of the Crypto Wars. It aims
to place current discussions in context. While it is sometimes claimed
that the conflict was settled in the 1990s, what emerges from this
overview is that a full articulation of privacy was never really
offered.

\subsection{Crypto Wars I: The Clipper Chip}\label{sec-clipper}

At the center of what has become known as the first Crypto Wars was the
Clipper Chip, which implemented the NSA-designed Skipjack algorithm for
an escrowed encryption system in encrypted telephony
\citep[p.~454]{Denning.1994ke}.\footnote{A different piece of hardware,
  the Capstone Chip, implemented Skipjack for computer communications
  \citep[p.~60]{Blaze.1994pf}.}

Clipper was the NSA's reaction to the announcement that AT\&T was about
to market the \$1,295 Telephone Security Device Model 3600 (TSD-3600)
\citetext{\citealp[p.~309]{Schneier.1997fb}; \citealp[p.~233]{Diffie:2010vm}}.
The news so worried the NSA that it briefed the President-Elect
Clinton's team and managed to make it a priority in the new
administration's first hundred days, with Vice President Gore leading
policy \citep[pp.~488--90]{EPP.1997bp}. The FBI, alerted by the NSA
\citep[p.~83]{Diffie:2010vm}, internally argued that warrant-resistant
communications were illegal \citep[pp.~308--309]{Schneier.1997fb}, and
even prepared a memorandum to AT\&T citing civil and criminal sanctions
\citep[pp.~309--310]{Schneier.1997fb}. The FBI was willing to go much
further, but the NSA's approach prevailed.

The government's strategy was to make adoption voluntary, but
inescapable. It planned to rely on the government's purchasing power
\citep[p.~32]{Froomkin:1996vm} and Fort Meade's secret trove of
cryptographic research to make its Escrowed Encryption Standard (EES) an
irrefutable proposition.

EES \citep{Denning.1994ke} used tamper-resistant hardware (Clipper and
Capstone) to encrypt 80-bit session keys with the NSA-designed Skipjack
cipher. Communication would be established, however, after the exchange
of a ``Law Enforcement Access Field'', LEAF, containing the session key
and a chip unique identifier, which were encrypted with a family key
common to all chips, as well as a chip unique key. Such keys were split
into components that were held by government agencies. Upon receiving a
legitimate request, such agencies would use the chip unique identifier
(obtained by the requesting authority by decrypting the LEAF with the
family key) to provide the chip unique key. Regulations designated NIST
and a division within the Treasury Department as the escrow agents
\citep[p.~759]{Froomkin:1995uz}.

Nine thousand TSD-3600E devices\footnote{The ``E''-designation models
  were equipped with the Clipper Chip
  \citep[pp.~318--319]{Blaze.2011ke}.} were acquired just by the
Department of Justice \citep[p.~769]{Froomkin:1995uz}, at the cost of
about \$1,000 per unit \citep[pp.~318--319]{Blaze.2011ke}.\footnote{The
  Clipper Chip itself cost between \$15 and \$25 per unit
  \citetext{\citealp[p.~480]{Dakoff:1996wz}; \citealp[p.~212]{NRC.1996}}.}
Mass government purchasing was thought to create market conditions
favorable to EES and unfavorable to alternatives
\citep[p.~442]{Magal.2003cp}. The NSA-designed Skipjack algorithm also
made superior security claims: it used a longer key (80 bits) and
deployed techniques only known to the Agency. The secrecy in the design
invited speculation that Skipjack hid deliberate vulnerabilities, which
the government sought to dispel by having it audited by a group of
external experts whose report corroborated the security claims
\citep[p.~126]{SKIPJACK.1995cl}.\footnote{Not everyone was satisfied:
  Schneier and Banisar argued the report's authors were too close to the
  NSA \citep[p.~315]{Schneier.1997fb}.} With this combination of
economic incentives and NSA security guarantees, the government hoped
EES would become the \emph{de facto} national standard
\citep[p.~442]{Magal.2003cp}, without any new statutory authority.

Yet escrowed encryption itself was challenged vigorously as
fundamentally unsafe. A report \citep{Abelson.1997} was particularly
influential; its authors, joined by other collaborators, have since
published two other reports that form the canon of the debate on
encryption policy \citep{Abelson:2015bl, Abelson.2024bo}. Despite this
opposition, Clipper was brought down not from the revelation of a
security flaw that exposed encrypted data (which was never found
\citep[p.~63]{NationalAcademiesofSciencesEngineeringandMedicine:2018br}),
though a vulnerability that allowed users to defeat the escrow itself by
spoofing the LEAF \citep{Blaze.1994pf} made front-page news in the June
2, 1994, issue of the \emph{New York Times} \citep{Markoff.1994fd}.

Rather, strong, organized opposition to EES coalesced around the costs
to technological innovation and a nascent digital economy (part of the
Clinton administration's information superhighway agenda, also
championed by Vice President Gore \citep[pp.~2--3]{Gurak.1998}). One of
the most important early responses was an open letter organized by the
organization Computer Professionals for Social Responsibility (CPSR), an
online petition to President Bill Clinton that Gurak considers
unparalleled at the time \citep[p.~41]{Gurak.1998}, signed by more than
50,000 people \citep[p.~7]{Kehl:2015vz}. It pointed to both privacy and
innovation: ``{[}I{]}f this proposal and the associated standards go
forward, even on a voluntary basis, privacy protection will be
diminished, innovation will be slowed'' \citep{CPSR.1994cc}.

The hardware-based approach that EES adopted was seen as particularly
hampering innovation, not least because it meant that security product
offerings would depend on government-authorized component suppliers
\citep[p.~237]{Diffie:2010vm}. Despite the relatively low cost per unit
of the chips, encryption ``was, by the 1990's, becoming an essentially
zero-marginal-cost technology, something that could often be implemented
in software more easily than by adding specialized hardware''
\citep{Blaze.2011ke}. Trying to salvage the plan, the government
switched to a software-based approach (which it named ``Software Key
Escrow'' \citep[pp.~33--34]{Froomkin:1996vm} but opponents dubbed
``Clipper II'' \citep[p.~320]{Schneier.1997fb}), partly in response to
congressional pressure against the plan
\citep[pp.~264--268]{Levy:2002wp}.

Yet there was still a more fundamental obstacle escrowed encryption
posed to the U.S. security industry. It believed the scheme would hurt
the U.S. global competitiveness \citep[p.~34]{Gurak.1998}. Companies
expected that their international clients were unlikely to buy a product
that gave the U.S. government, but not their own governments, access to
their encrypted data. This issue was folded into a broader push by the
industry for loosening export controls imposed through Arms Export
Control Act and Export Administration Act authorities and implemented by
the International Traffic in Arms Regulations (ITAR), which limited
exports to 40-bit keys \citep{Diffie.2007ec}.\footnote{While less
  relevant to this article, export controls were a significant part of
  Crypto Wars I. The Export Administration Act also regulated the export
  of dual-use (military and non-military) products
  \citep[p.~114]{NRC.1996}. A complex set of rules and individual
  evaluations established what could be exported
  \citep[pp.~21--23]{Froomkin:1996vm}; in practice, systems with keys
  longer than 40 \emph{bits} were not allowed. The legislation also
  allowed for the regulation of \emph{imports}, but a 1996 National
  Research Council report noted that the regulations imposed no
  restrictions \citep[p.~115]{NRC.1996}. Practical reasons nonetheless
  often imposed the development of a single product for both domestic
  and international markets
  \citetext{\citealp[p.~277]{Banisar:1999tr}; \citealp[p.~262]{Levy:2002wp}}
  or the adoption of a single global consumer standard
  \citetext{\citealp[p.~326]{Schneier.1997fb}; \citealp[p.~158]{NRC.1991cr}}.
  As such, the same NRC report concluded export controls were used
  indirectly to set domestic encryption restrictions
  \citetext{\citealp[pp.~113--114]{NRC.1996}; \citealp[see
  also][p.~357]{Froomkin:2006ty}}.} Like the strict controls on
encryption, which U.S. exporters complained hurt their global standing
\citep[p.~729]{Diffie.2007ec}, Clipper would put their products at an
international disadvantage.

On a parallel track, the U.S. had sought support for an international
solution to soothe industry concerns. It tried to inscribe key escrow
requirements into the non-binding 1996 Wassenaar Arrangement (which, at
the time, limited exports to 64-bit symmetric keys\footnote{The
  Wassenaar Arrangement system, which lets each country decide on its
  own restrictions \citep[p.~732]{Diffie.2007ec}, loosened on encryption
  in 2009 \citep[p.~60]{Gill:2018us}. It has since become broadly
  permissive \citep[pp.~3--4]{Upson.2021}.}). That effort failed
\citep[p.~784]{Charlesworth.2007pe}. A different effort was made through
the OECD. Only France and the U.K. backed the proposal. The document the
organization ultimately published\footnote{Recommendation of the Council
  concerning Guidelines for Cryptography Policy. OECD/LEGAL/0289.} did
not adopt the key-escrow strategy proposed by the U.S. Though it called
for reconciling market-driven technological development and
investigative demands, it was vague and let members free to decide for
themselves \citep[p.~100]{Koops:1999uk}. The global Clipper ``effort
failed'' \citep[p.~449]{Magal.2003cp}.

A congressionally commissioned report by the National Research Council
(NRC) had also dealt a setback to the Clipper Chip \citep{NRC.1996}.
Although Clipper critics had been wary that the NRC-convened group was
dominated by members who were or had been part of the intelligence and
law enforcement communities to deliver an independent assessment
\citetext{\citealp[pp.~242--243]{Diffie:2010vm}; \citealp[p.~296]{Levy:2002wp}},
the report criticized key parts of the government's strategy. The
report's take-home message was telegraphed by its title,
\emph{Cryptography's Role in Securing the Information Society} (CRISIS).
U.S. policy was ``not adequate to support the information security
requirements of an information society,'' the group concluded
\citep[p.~301]{NRC.1996}.

\subsection{Crypto Wars II: ``Going Dark'' and ``Apple v.
FBI''}\label{sec-applevfbi}

The doomsday premonitions by 1990s officials about widespread encryption
were exaggerated. As DeNardis writes, ``those who developed encryption
standards in the 1970s, whether for securing financial data or for
preserving the confidentiality of government communication, could never
have predicted that cryptography would not be ubiquitously deployed
everywhere well into the twenty-first century''
\citetext{\citealp[p.~218]{DeNardis:2020ts}; \citealp[pp.~257--259]{Diffie:2010vm}}.
As late as 2013, Narayanan distinguished between the success in the
deployment of what he called ``crypto-for-security'' and the slow pace
of ``crypto-for-privacy'' \citep{Narayanan:2013ij}. Even
well-established encryption disappointed. Jarvis notes that, in 2016,
only 40\% of internet traffic was encrypted with Transport Layer
Security (TLS); the figure would reach 80\% in 2019
\citetext{\citealp[p.~338]{Jarvis.2021cw}; \citealp[see][ for a
discussion]{Kerschbaumer.2025sh}}.

In late 2014, when, in the wake of the Snowden revelations
\citep[pp.~19--27]{Landau.2013ms, Landau.2014hm, Anderson.2020}, Apple
\citep{NYT.2014ps}, Google \citep{WaPo.2014ae}, and Meta's WhatsApp
\citep{Wired.2014wa} announced they were implementing end-to-end
encryption by default, the 1990s anxieties seemed to finally
materialize. This meant that officials were no longer able to ``seek
access to stored communications held by these intermediaries by
obtaining a warrant, court order, or subpoena''
\citep[p.~4]{Zittrain:2016uv} as they had before.

The FBI's reaction to end-to-end encryption by default was quick. A
month after Apple's announcement, Director James Comey gave a speech at
the Brookings Institution claiming ``it would have very serious
consequences for law enforcement and national security agencies at all
levels'' \citep{FBI.2014gd}. He echoed FBI General Counsel Valerie
Caproni's 2011 warning that officials were ``going dark''
\citep[p.~7]{Caproni.2011gd}. To Comey, the phrase denoted the problem
that ``{[}t{]}hose charged with protecting our people aren't always able
to access the evidence we need to prosecute crime and prevent terrorism
even with lawful authority'' \citep{FBI.2014gd}.

The conflict was not limited to the U.S. For instance, in 2015,
Brazilian courts ordered the suspension of WhatsApp after it deployed
end-to-end encryption (E2EE) \citep{Abreu:2018ga}. The Supreme Court
stayed the order, but the constitutional challenge (ADPF 403) is still
pending \citep{Silva.2021eb}. But Crypto Wars II became known for the
orders in U.S. cases, often informally referred to as the ``Apple v.
FBI'' cases.

The case pursued by U.S. law enforcement officials before Congress and
the court of public opinion \citep{Caproni.2011gd, FBI.2014gd} elicited
responses
\citep{Bellovin:I2sir3KP, Zittrain:2016uv, Swire.2011eg, Kehl:2015vz, McConnell:2015vu, UnitedNationsHumanRightsCouncil:2015ta}.
The Berkman Klein Center at Harvard University published an influential
report \citep{Zittrain:2016uv}, and authors of a 1997 report on Crypto
Wars I proposals \citep{Abelson.1997} were joined by others, and
published the leading analysis ``Keys under doormats''
\citep{Abelson:2015bl}.

The conflict centered on requests for Apple to assist law enforcement
officials in overcoming security features in its own systems. In 2015,
the FBI applied for orders directing Apple to assist with investigations
by disabling a security feature that erased the device's data after a
number of failed passcode entry attempts
\citep{Pfefferkorn:2018uf, Rozenshtein:2018wq}. This barred the FBI from
launching a brute-force attack against devices it had in its possession
and which it was authorized to access. In one of the cases, after a
terrorist attack that took the lives of 14 victims in San Bernardino,
CA, the FBI obtained permission to access the iPhone used by one of the
shooters from the local government agency that employed him and owned
the phone \citep{NYT.2020ae}. In that case and another in Brooklyn, the
government applied for an All Writs Act order compelling Apple to design
and use its cryptographic keys to sign and install a modified version of
iOS without the passcode-entry deletion trigger
\citep[pp.~125--128]{Rozenshtein:2018wq}.

Apple responded, contesting the government in an open letter to its
customers, signed by its CEO \citep{Cook.2016mo, NYT.2020ae} and in
court. Its arguments in court primarily were that the government's
orders either were not within the ambit of the All Writs Act or did not
meet the tests established by case law for similar assistance. The
company also argued that, as a constitutional matter, compelling it to
comply with the order would be too burdensome and a violation of the
Fifth Amendment's due process clause, and it had a First Amendment right
against compelled speech that protected it from writing the code for the
modified version of iOS and signing it with its cryptographic key
\citep[pp.~125--128]{Rozenshtein:2018wq}.

The clash was mooted when the FBI announced it had been able to unlock
the phones with help from a third party \citep{NYT.2016ap}.
Section~\ref{sec-protocols-device} analyzes the security considerations
raised by commentators \citep{Abelson:2015bl, Zittrain:2016uv} as it
discusses proposals put forward since then.

\subsection{Crypto Wars III: A New Chapter?}\label{sec-cryptowarsiii}

Chronicling the Clipper Chip debacle and the loosening of export
controls in late 2000, Steven Levy crowned a clear winner: ``It was
official: public crypto was our friend'' \citep[p.~307]{Levy:2002wp}
Indeed, writing about the ``Apple v. FBI'' cases in 2016, he questioned
why the Crypto Wars were being rehashed. To Levy, the conflict had been
settled, and the deal was that, if the government ``wanted to gain
access to encrypted communications and files, they would do so by
warrants and their own cryptanalysis, and not by demanding that the
systems themselves should be weakened'' \citep{Levy.2016wa}. Rozenshtein
correctly countered that public policy is not subject to stare decisis
\citep[p.~1195]{Rozenshtein:2019vf}.

Not only is there no \emph{res judicata}, but Levy's 2000 conclusion was
premature. Snowden revelations commentators believed that a mass
surveillance scheme was made possible by a deliberate vulnerability they
suspected had been introduced by the NSA to Dual EC DRBG, a pseudorandom
number generator
\citep[pp.~324--327]{Perlroth.2013na, Bernstein.2016ec, Weippi.2016, Jarvis.2021cw},
which had the potential of exposing all TLS/SSL encrypted internet
traffic. And in at least one case before the Apple cases, the FBI sought
and obtained an order compelling an email provider to hand over its
private SSL certificate keys so that the police could launch a
man-in-the-middle attack targeting Edward Snowden
\citep{Levison.2014lb, Owsley.2017lb, Pfefferkorn:2017vd}. Accounting
for this, the dividing line between the Clipper Chip saga and ``Apple v.
FBI'' seems less clear.\footnote{``One might say that when Crypto War II
  ended, the NSA took the fight underground, a move entirely unknown to
  the public until Edward Snowden'' \citep[p.~288]{Anderson.2021re}.}

Likewise, the 2015--2016 skirmishes seemed to quickly give way to a
truce \citep[p.~900]{Koops:2018bh}. Yet not much later, as we will see
(Section~\ref{sec-apple-csam}), Apple reignited the debate when it
announced updates introducing client-side scanning for child sexual
abuse material, which commentators speculated owed at least in part to
pressure from officials against plans to adopt E2EE for iCloud data
\citep{Reuters.2020af}. Not long after, the European Union started
considering client-side scanning as part of its proposed Child Sexual
Abuse regulation, which is informally referred to as ``Chat
Control.''\footnote{Proposal for a Regulation of the European Parliament
  and of the Council laying down rules to prevent and combat child
  sexual abuse, COM(2022) 209 final, 2022/0155(COD). \emph{See}
  \citep{Tuchtfeld.2022ty, Backer.2022ms}.}

As of this writing, the EU proposal had been softened to remove language
making client-side scanning mandatory \citep{Clark.2025er}, though some
have concerns that other provisions could be used to reintroduce the
requirement in the implementation of the regulation. Apple eventually
rolled out E2EE for iCloud data, which it calls Advanced Data
Protection; in the U.K., officials publicly opposed it and later issued
an order that it be disabled \citep{Menn.2025uo}. Apple disabled
Advanced Data Protection in the U.K. and challenged the order, a
technical capability notice under the Investigatory Powers Act of 2016;
the government sought but was rebuffed in its effort to keep the
challenge under seal \citep[pp.~522--523]{Keenan.2025ii}. It was
initially reported that British authorities had agreed to drop the order
after U.S. high-ranking officials (including the President, the
Vice-President, and the Director of National Intelligence) intervened
\citep{Miller.2025uh}. Later, Apple stated it was ``still unable'' to
offer Advanced Data Protection in the U.K., and news accounts said a
second order had been issued, which ``stipulated that the order applied
only to British citizens' data'' \citep{Gross.2025um}. The status of
encryption regulation remains uncertain in other jurisdictions, as noted
above.

One way to understand this is as a new chapter of the Crypto Wars, its
third installment. Jarvis \citep{Jarvis.2021cw} argues that a proper
account of the conflict would begin much earlier than the Clipper Chip,
with the NSA's reaction to Kahn's \emph{Codebreakers}.\footnote{Other
  periodizations have been offered. Landau suggests that ``{[}t{]}he
  first skirmish in the `Second' Crypto War occurred in 2003 when the
  FBI recommended extending the CALEA to Voice over IP (VoIP)
  providers.'' \citep[p.~242]{Landau.2022dc}. On CALEA, see
  \citep{Hurwitz.2016ec}.} He counts ``Apple v. FBI'' as the third act.
The fact that the push to limit user encryption now comes not just from
government officials but also from companies that were mass adopters of
E2EE might be a distinguishing factor. And while part of the debate is
still about law enforcement, part is not, as Duan and Grimmelmann note
\citep[p.~15]{Duan.2024cm} in their comprehensive analysis of content
moderation on E2EE messaging services under U.S. statutory law.

I make no periodization claims. My interest, instead, is to show that,
despite longstanding debates and occasional premature declarations of
victory, a central question has not been addressed. Others have noted
open questions. Feigenbaum observes that ``it has not actually been
shown that no useful form of LEA can be implemented without creating
unacceptable risk'' \citeyearpar[p.~29]{Feigenbaum:2019fy}. Hurwitz
\citeyearpar[pp.~369--371]{Hurwitz.2016ec} sees a different social,
economic, and technological landscape posing new questions that cannot
be taken as settled by the debates surrounding the Clipper Chip.
Rozenshtein goes further and submits that ``{[}t{]}here can and will be
no permanent resolution to the problem of law enforcement access to
encrypted data'' \citep[p.~1196]{Rozenshtein:2019vf}. My claim is that,
despite the sophistication the debate has acquired, our conceptual
understanding of how privacy is implicated remains limited.

\section{Securing Privacy}\label{sec-securing}

As DeNardis observes, arguments about technological innovation and its
economic promises, which weighed heavily in the resolution of the Crypto
Wars I, if present, are much less conspicuous today
\citep[p.~126]{DeNardis:2020ts}. Neither do privacy advocates turn to
libertarian cypherpunk arguments that animated some of the Crypto Wars I
\citep{Queiroz.2022, Jarvis.2021}. Instead of a dispute over the
authority of the state to impose restrictions on encryption, what
defines the discourse in opposition to the FBI and officials in the
United States, Brazil, Australia
\citep{Hardy.2020, Stilgherrian.2019ea, Stilgherrian.2021ea, Davis.2022da},
India
\citep{Ahmad:2009ce, Mohanty.2019ei, Mohanty.2021ei, Burman.2021ei} and
in the U.K. \citep{OTI.2017uk, Keenan.2020uk} is a savvy \emph{reframing
of the terms of the conflict}.

While officials claimed that the security of citizens depended on a
sacrifice of their privacy and justified limiting encryption
\citep{Baker.1994dw, Comey.2016ep, Rosenstein.2017ec}, the other side of
this debate questioned the very premise of this opposition: it was not
privacy versus security, but security versus security
\citep[pp.~617--619]{Pell:2016wm}. To be fair, the 1996 NRC report also
argued that viewing cryptography as an opposition between security and
privacy was ``simplistic'' \citep[p.~302]{NRC.1996}:

\begin{quote}
If cryptography can protect the trade secrets and proprietary
information of businesses and thereby reduce economic espionage (which
it can), it also supports in a most important manner the job of law
enforcement. If cryptography can help protect nationally critical
information systems and networks against unauthorized penetration (which
it can), it also supports the national security of the United States.
\end{quote}

Yet what the reports' authors considered in a sage exercise of
futurecasting is now our rich reality. And scholars have done important
work in articulating it. Susan Landau, a leading voice in encryption
policy who has written on it for almost three decades
\citetext{\citealp[p.~286]{Anderson.2021re}; \citealp{AMS.2023la}},
deserves much credit for reframing security as an ally of privacy and
the use of encryption. Though earlier writing gestured toward this move
\citep[pp.~11--12]{Bellovin.2006si, Landau:2011wy}, its current form
appears in her testimony at a 2016 public hearing before the United
States House of Representatives \citep[p.~24]{Landau.2016et} and is
articulated in monograph-length treatment \citep{Landau:2017uy}.

In that book \citeyearpar{Landau:2017uy}, Landau places the discussion
within a broader cybersecurity perspective. She notes how fragile
systems security is and shows how digitization has left everyone exposed
to its fragility, from the smartphone user accessing her bank to a
nation's critical infrastructure. She describes how vulnerabilities
become weapons for cyberwarfare, as in the U.S. attack on Iran's nuclear
program. Against this backdrop, she concludes that exceptional-access
schemes like those proposed by the FBI exact a steep security price in
exchange for evidentiary demands: ``The government's role is to provide
security---national security and law enforcement---and not to prevent
individuals from maintaining their own security''
\citep[p.~172]{Landau:2017uy}.

This reframing of the conflict is insightful. It removes the
``home-field advantage'' that law enforcement and intelligence officials
enjoy when the issue is cast as privacy \emph{vs.} security
\citep[p.~41]{Solove:2011vz}. Privacy usually plays away games on the
fields where security questions are decided. It is not only that the
institutional design of those fora is oriented toward security as a
paramount value. Privacy advocates do not even have enough information
to contest the authorities, who benefit from the secrecy imposed on
matters of national security and (albeit to a lesser extent) public
safety \citep[p.~772]{Solove:2007up}.

By treating encryption restrictions as a conflict internal to security,
privacy advocates ``turn{[}{]} security arguments on their heads''
\citep[p.~207]{Moore:2010tm}. They reverse how encryption was treated
for most of its trajectory, until the turn in the 1970s: once discussed
as a matter of national security and regulated on reasons of state,
encryption is now cast as a matter of the security of each individual in
a society, not only of the state as that society's protector. The right
to privacy as a right to security speaks in the language that once
belonged only to government officials, even when it relies on ominous
scenarios and catastrophe, like ``a digital Pearl Harbor''
\citep[pp.~126--127]{DeNardis:2020ts}.

Much of this argument in encryption policy was advanced by cryptography
and systems security scholars, including in policy writing and
congressional briefing, e.g.
\citep{Landau.2016et, Abelson:2015bl, Abelson.2024bo}. Other influential
writing has featured scholars from law and related fields
\citep{Zittrain:2016uv, Pfefferkorn:2018uf}. This important body of work
has set aside for strategic reasons (sometimes explicitly,
\citep[p.~24]{Zittrain:2016uv}) normative discussions that go beyond the
security framing.

Privacy-as-security proponents are in good company. In fact, encryption
policy has developed in tandem with leading scholarship on privacy,
which has been marked by pragmatic thinking
\citetext{\citealp[e.g.][]{Solove:2010vr}; \citealp{Hartzog.2021wp}; \citealp[
p.~77]{Gutwirth:2006up}; \citealp[see][ pp.~216--217]{APLM.2023pc}}.
This scholarship emphasizes procedural questions of oversight \citep[
p.~1529]{Solove:2010vr}, and (though perhaps this is not as widely
shared among the field \citep[ pp.~67--68]{Richards.2022wp}) rejects
non-consequentialist accounts of its value \citetext{\citealp[
pp.~1144--1145]{Solove:2002gr}; \citealp[ p.~88]{Solove.2008up}}. It
should not surprise, then, that encryption policy has likewise focused
on such questions.

In fact, while U.S. courts have yet to confront fundamental rights
claims about encryption and surveillance, European courts have---and
have embraced the pragmatic approach to privacy that emphasizes
procedural questions and is amenable to the security conception of
privacy at play in encryption policy. In \emph{Podchasov v. Russia}
\citep{Podchasov.2024hr}, the European Court of Human Rights ruled that
providers could not be required to build systems to maintain decryption
capabilities. It seized on the impacts on the safety of all users to
conclude that the Russian requirement was disproportionate
\citep{Basri.2025pr}. A commentator writes that ``the \emph{Podchasov}
judgment seems to leave options open if a decryption technology becomes
available that would not weaken the security of all users \citep[
p.~5]{Shurson.2024er}. In previous judgments, both the ECtHR and the
Court of Justice of the European Union, applying different legal
regimes, emphasized procedural safeguards (or the lack thereof) \citep[
p.~2]{Zalnieriute.2022bb, Basri.2025pr}.\footnote{This is by no means a
  full analysis of either applicable EU law or Council of Europe law.
  \emph{See}
  \citep{Zalnieriute.2022bb, Davis.2024re, Shurson.2024er, Basri.2025pr}.
  Nor is the claim here that this emerging case law has reduced privacy
  to safety and oversight; rather, these have been the most salient and
  more developed discussions. For an argument that exceptional-access
  mandates are inconsistent with both European regimes because they
  infringe upon the ``essence'' of the right to privacy, see \citep[
  pp.~68--72]{Davis.2024re}.} Procedural safeguards are part and parcel
of the pragmatic approach to legal thinking on privacy, as we have just
seen. And, as we shall see next, most proposed lawful-surveillance
protocols are anchored in procedural safeguards. Both scholarly writing
and judicial doctrine have articulated privacy in a manner that is
consistent with its treatment in encryption policy.

Securing privacy has been a strategic and important move. My goal is not
to refute it. Yet it does not fully capture what is at stake. I make
this case in the remainder of the article.
Section~\ref{sec-surveillance-protocols} first surveys and examines
vigorous work, which I gather under the heading of lawful-surveillance
protocols. This agenda, I argue, fits the mold of privacy as security
and seeks to meet the goalposts it sets. Such work necessitates
conceptualizing privacy to account for more than security.

I should stress that my argument is \emph{not} that cryptography and
systems security scholars ignore or do not value privacy.\footnote{Landau,
  for instance, while focusing on security, also refers to the preamble
  to the U.S. Constitution and the origins of the Fourth Amendment, and
  recalls the Church Committee \citetext{\citealp[ pp.~189--190,
  p.~250]{Landau:2011wy}; \citealp[ p.~169--172]{Landau:2017uy}}.} My
claim is also not that any or all of those scholars subscribe to a
strong version of privacy as security, according to which all important
aspects of the encryption debate are reduced to security, broadly
construed. Rather, my position is that privacy-as-security has been the
only articulation to gain traction.\footnote{Of course, I do not ignore
  that advocacy will often appeal to the value of privacy, nor do I
  question its worth. This just does not provide the conceptual
  articulation I argue is missing.}

Next, Section~\ref{sec-beyond-security} offers two reasons why
privacy-as-security does not explain the stances that encryption
restriction opponents have taken, even when lawful-surveillance
protocols are bracketed. It also examines how an influential critique of
client-side scanning carries within it the seeds of concerns that exceed
security.

\section{Lawful-Surveillance
Protocols}\label{sec-surveillance-protocols}

As FBI officials spoke of ``going dark'' in testimony before Congress,
some computer scientists began thinking about lawful-surveillance
protocols \citep{Segal:2016bh}. Pioneering work on this agenda
positioned itself as a pragmatic response to Edward Snowden's
revelations \citep{Feigenbaum:2014el}. This response is premised on the
idea that surveillance carried out by intelligence and police agencies
can proceed through open processes, using cryptographic protocols to
allow the public to oversee their activities, rather than secret
processes that require society to trust these agencies without concrete
information on which to rely. Proponents of lawful-surveillance
protocols regarded surveillance as an inevitable reality
\citetext{\citealp[ p.~1]{Segal:2014tl}; \citealp[
p.~3]{Feigenbaum:2017ur}} and argued that the best way for cryptography
and information-security research to control it is ``to design protocols
that allow government agencies to collect and use data that are
demonstrably relevant to their missions while respecting the privacy of
ordinary citizens and being democratically accountable''
\citep[p.~1]{Segal:2014tl}.

Lawful-surveillance protocols posit that ``by deploying appropriate
cryptographic protocols in the context of sound policy and the rule of
law, citizens can have both user privacy and effective law enforcement
and intelligence'' \citep[p.~1]{Feigenbaum:2017ur}. Their thinking was
attuned to influential scholarly writing by Solove, to whom privacy and
security can be reconciled ``by placing security programs under
oversight, limiting future uses of personal data, and ensuring that the
programs are carried out in a balanced and controlled manner''
\citep[p.~207]{Solove:2011vz}. The agenda follows a broader trend of
proceduralization in data protection \citep{Gutwirth:2006up}.

Part of this project requires establishing the design principles for the
lawful-surveillance protocols so that ``citizens can have both user
privacy and effective law enforcement and intelligence''. The goal is to
confer upon the qualifier in ``lawful surveillance'' a stronger meaning,
not just an indication of the agent behind it (as in ``government
surveillance''). The agenda aims to establish ``lawful, accountable,
privacy-preserving surveillance. The idea is to combine cryptographic
protocols (SMC, PIR, etc.) with blackletter law'' \citep[
p.~10]{Feigenbaum:2017hg}. Proponents ``advocate{[}{]} combining
technical protocols with legal and social protocols'' \citep[
p.~10]{Feigenbaum:2017hg}.

This section surveys work that, even if it does not label itself as
such, is helpfully understood from the perspective of
lawful-surveillance protocols.

\subsection{Data Available in Cleartext}\label{sec-protocols-cleartext}

The first protocols were aimed at cryptographically regulating access to
data that government officials already enjoyed in cleartext
\citep[p.~11]{Feigenbaum:2017hg}, such as data stored by cloud service
providers \citep{Feigenbaum:2014el}.

A real case motivated a protocol concerning cell towers, the ``high
country bandits'' case \citep[ pp.~805--807]{Henderson.2013rh}. In 2010,
a group robbed banks in three different cities in Arizona and Colorado.
A witness testified to seeing members of the group talking on a cell
phone. Based on that lead, the FBI obtained cell-tower connection
information for the crime locations and used it to identify cell phone
numbers that appeared on all three lists. This allowed officers to
single out a suspect, request his subscriber information (name and
address) from the phone carrier, and arrest him.

However, in order to compare the cell-tower connection lists and find
phones that appeared on all three, the metadata of 149,999 people was
disclosed to the FBI, making them vulnerable to use in other
circumstances and for other purposes
\citep[pp.~11--12]{Feigenbaum:2017hg}. Instead, the academics proposed a
protocol that would allow the phone carrier to keep information about
connections to cell towers \emph{encrypted} while still meeting the
FBI's demand for numbers of potential suspects that recur across the
different \emph{datasets}. The information corresponding to potential
suspects (i.e., those who appeared on all three lists) would be
delivered to the FBI without exposing the hundreds of other records; the
FBI would have access only to the results of that query across the
records. The carrier itself would also not know what or whom the FBI's
interest targeted, because the query (i.e., the specification of the
parameters sought by the FBI: phones connected to towers in the vicinity
of the crimes at the times they were committed) would be encrypted.

\subsection{Data Stored in Encrypted
Devices}\label{sec-protocols-device}

Other proposals seek to create protocols that provide government
officials with cleartext access to data stored in encrypted devices. For
instance, data contained in smartphones with full-disk encryption (FDE),
which protects data stored on a powered-off device
\citep{Zinkus.2022cc, Fruhwirth.2005de}, and run-time file encryption
(RTFE), which protects devices that are powered on but locked
\citep{Zinkus.2022cc}.\footnote{See \citep{Zinkus.2022ds} a description
  of the security mechanisms in iPhone and Android devices.}

\subsubsection{Self-Escrow}\label{sec-savage-ozzie}

Stefen Savage's proposal identifies two main considerations in debates
about encryption regulation. First, the risk of mass surveillance.
Second, the risk that an exceptional-access mechanism would be exploited
by malicious actors \citep[p.1764]{Savage:2018cd}. To address these two
points, his proposal lists four properties of a suitable system for
granting access to authorities:~\emph{non-scalability}, with constraints
that prevent the system from being used for mass
surveillance;~\emph{authorization}, with verification of the legal basis
for access;~\emph{particularity}, via conditions that allow the
mechanism to provide access to a single device only;
and~\emph{transparency}, through conspicuous notice to the user when
their device is the target of exceptional access
\citep[p.1765]{Savage:2018cd}.

Under his protocol, exceptional access requires that the device be
physically obtained and held for a period before which the mechanism
would not operate, as a way to avoid mass surveillance and to make
clandestine interventions on the device more difficult (e.g., taking
advantage of a user's inattention to compromise the device). The key
would be self-escrowed on the device itself to avoid the risks of
centrally managing millions of escrowed keys. To provide transparency,
the device would display a message upon boot in the event of an unlock
attempt. The mechanism would also require the manufacturer's
cooperation, which, according to the author, adds an external check on
the police by an entity that already enjoys the user's trust.

Savage's proposal partly follows principles adopted by Ray Ozzie in his
proposal, which gained attention with a \emph{Wired} article by Steven
Levy \citep{Levy.2018cr}. Ozzie did not publish a paper outlining his
proposal, ``CLEAR,'' which was discussed in confidential, informal
conversations. After the magazine story, Ozzie published slides related
to the proposal \citep{Ozzie.2018cr}. The scheme would also be based on
self-escrow and unlocking via a key provided by the manufacturer. But
Ozzie did not include Savage's time vaulting and transparency
components.

Ozzie's proposal was heavily criticized
\citep{Bellovin.2018tr, Landau.2018ro}. Upon seeing it presented
\citep{Bellovin.2018sf}, Eran Tromer immediately identified how an
adversary could manipulate the police by pretending to target a
suspect's device in order to obtain the key to anyone else's device
\citep{Tromer.2018cr}. Tromer's attack exploits the fact that Ozzie's
protocol does not verify whether the QR code displayed by the device
seized by the police actually corresponds to that device. This would
create room for someone to hand over to the police a device that
impersonates another device, displaying the QR code for someone else's
device (for example, a victim's phone stolen by the criminal). When the
police entered the unlock code into the seized device, that device would
then send that information to the attacker (who could thus unlock the
victim's phone). Savage outlined potential mitigations for this attack
but did not present a protocol that integrated them
\citep[p.~1774]{Savage:2018cd}.

An important factor in Savage's and Ozzie's proposals is control over
the manufacturer-held authorization or decryption keys required to
access the passcode self-escrowed on the device once police had
physically obtained it. This would create an opportunity for malicious
insider attacks---people working for manufacturers managing those
keys---as well as for external attacks aimed at obtaining those keys,
including via social engineering. This has been identified as a risk of
key-escrow schemes since the 1997 ``Risks'' report
\citep[pp.~11--12]{Abelson.1997} and reiterated in ``Keys under
doormats'' \citep[p.~4]{Abelson:2015bl}. A consequence of key-escrow
proposals is the creation of new, highly valuable assets for
cybercrime---vaults holding billions of cryptographic keys.

Savage's and Ozzie's proposals partly mitigate this risk by limiting the
immediate utility of compromising manufacturers' keys, which would not
give direct access to the contents of encrypted devices (which, again,
would require physical seizure). Savage compares these risks to those
inherent in managing the cryptographic keys manufacturers use to sign
operating-system updates as legitimate for devices.

Matthew Green points out that the comparison is not apt, because
operating-system update keys are used far less frequently---only when
updates are issued, for example, monthly
\citep[p.~60]{Green.2018ro, NationalAcademiesofSciencesEngineeringandMedicine:2018br}.
This higher frequency of use for a decryption key would also imply that
more people have access to it to handle numerous requests throughout the
day, as noted by Pfefferkorn
\citetext{\citealp[p.~7]{Pfefferkorn:2018uf}; \citealp[p.~60]{NationalAcademiesofSciencesEngineeringandMedicine:2018br}}.
A National Academies of Sciences, Engineering, and Medicine (NAS) report
further stresses that compromising an unlock key would give immediate
access to a device's contents, yielding more direct utility to an
attacker than an update-signing key, which would merely open an
opportunity for launching an attack
\citeyearpar[pp.~59--60]{NationalAcademiesofSciencesEngineeringandMedicine:2018br}.

Savage responded that abuse risks could be mitigated with a
key-protection mechanism that requires a certain number of people for
access, with auditability via access reports
\citep[p.~1770]{Savage:2018cd}. As further mitigation, Varia suggests
that exceptional access could occur on the same cadence as current
software updates \citep{Varia.2018le}.

\subsubsection{Social Cost}\label{social-cost}

Another proposal follows the properties identified by Savage for a
lawful-surveillance system but modifies them with respect to
authorization and transparency. The principle behind the JJE
system---\emph{judge, jury and encryptioner}---is the imposition of a
``social cost'' for decryption \citep{ServanSchreiber:2020ts}. Unlike
Savage's and Ozzie's proposals, JJE does not assign the manufacturer the
function of unlocking the device. The manufacturer merely develops the
system and produces the device in accordance with it. Unlocking then
depends on the involvement of two groups: custodians and delegates. The
former would be entities ``belonging and operated by the government
itself (e.g., district courts), nonprofit groups, corporations, or
academic institutions'' \citep[p.~5]{ServanSchreiber:2020ts}. Custodians
would jointly (in a number set by the system) provide an unlock request,
which would depend on the approval of the delegates. With that request,
the police would then need to find the delegates---the
``jurors''---third-party devices whose physical possession would be
required to approve the unlock. These delegates would be randomly chosen
and known only when the unlock request was approved.

The authors admit that their protocol would not protect devices in
situations where ``abusive law enforcement to secretively coerce all
custodians or to routinely silence citizens selected as delegates''
\citep[p.~13]{ServanSchreiber:2020ts}. This would also extend to
attackers external to the police, such as criminals, even if they would
have to compromise custodians and obtain the jurors' devices.

\subsection{Encrypted
Communications}\label{sec-protocols-communications}

Designing lawful-surveillance protocols is much harder with respect to
encrypted \emph{communications} (as compared with data \emph{stored} on
encrypted devices) \citep[p.~10]{EncryptionWorkingGroup:2019wr}. One
reason is that good security practice for communications is to adopt
\emph{forward secrecy}, in which a symmetric key for each session is
established and promptly discarded. The advantage of this approach is
that it limits any eventual interference by an attacker, who would have
access only to communications from the moment of the attack, as opposed
to what would happen if asymmetric keys were used to encrypt messages,
which would give the attacker access to all communications (that had
been encrypted with that key). While email-encryption systems such as
PGP use asymmetric keys to encrypt messages, apps such as WhatsApp and
Signal adopt \emph{forward secrecy} \citep[p.~6]{Abelson:2015bl}.
Moreover, whereas proposed lawful-surveillance protocols for stored data
rely on the requirement of physical possession of the device to limit
risk, authorities' demands for communications data generally translate
into remote access to communications data intercepted in transit. This
adds more complexity and risk to the system
\citep[p.~65]{NationalAcademiesofSciencesEngineeringandMedicine:2018br}.
A report by a working group composed of people with experience in
law-enforcement and intelligence agencies and independent academics
considered that research on exceptional access for stored data deserved
more discussion but did not ``identify any approach to increasing
law-enforcement access that seemed reasonably promising''
\citep[p.~8]{EncryptionWorkingGroup:2019wr}.

\subsubsection{Crumple Zones}\label{sec-crumple-zones}

One proposed lawful-surveillance protocol that would encompass encrypted
communications (and could also be applied to stored data) is described
as creating ``crypto crumple zones'' \citep{Wright:2018dj}. The term is
an allusion to the automotive engineering structure that allows a
vehicle to deform in a way that is safe for passengers in the event of a
collision: the car crumples in a controlled fashion, absorbing the
impact. Wright and Varia suggest that encryption could also ``break a
little bit in order to protect the integrity of the system as a whole
and the safety of its human users'' \citep[p.~289]{Wright:2018dj}.

The scheme embeds cryptographic puzzles into per-message ephemeral keys,
making recovery possible but very expensive. Part of the puzzle is
designed to match Bitcoin's proof-of-work computation so the authors can
predict real-world marginal cost with reasonable confidence
\citep[p.~289]{Wright:2018dj}. Rather than introducing an escrow key (or
escrowing key bits), this approach offers access by crumpling the
encryption key---planned and calibrated insecurity
\citep[p.~292]{Wright:2018dj}. Wright and Varia attribute the problem of
mass surveillance to a lack of proportionality in its costs, which their
crumple zones would undo \citep[p.~290]{Wright:2018dj}:

\begin{quote}
When surveillance is cheap, law enforcement agencies have little
incentive to refrain from collecting information on as many targets as
possible\ldots. When electronic surveillance is expensive, as in the
physical world, law enforcement must choose their targets wisely to
focus on only the most pressing threats.
\end{quote}

To that end, they estimate the computational cost for solving the
puzzles that would grant access to encrypted data. By setting an entry
cost on the order of hundreds of millions to billions of dollars, they
intend that such an investment will select only state agents, who would
have such sums to gain access to the data. Other costs would also be
set: per geographic region, application or system, user, and message.
Each break would require marginal costs (which they estimate between a
thousand and a million dollars) in addition to the entry costs, creating
economic incentives so that only legitimate targets would be pursued.

Although it depends on fewer external factors and reduces dependence on
trust in companies (compared to the escrow schemes above), this proposal
does not ensure transparency \citep[p.~3]{ServanSchreiber:2020ts} or
legal authorization for obtaining the data (i.e., it does not establish
a procedure by which only legitimate agents could decrypt the data).

\subsubsection{Arleas}\label{arleas}

A more recent proposal applied to encrypted communications,
ARLEAS---\emph{abuse resistant law enforcement access system}---also
explores blockchain concepts, but in its public ledger, not
proof-of-work \citep{Green.2021ar}. Messages in an ARLEAS system would
be encrypted both with the interlocutors' keys and with a second cipher
that enables exceptional access. Decryption via this second cipher
requires that the police obtain a judge's authorization and that this
authorization be validated in the system by a transparency function,
which publishes information about the interception in the public ledger.
Only with the proof of publication related to the authorization can the
message be decrypted using the exceptional-access cipher. As the name
indicates, the system is not abuse-proof; abuse would occur if judge and
police acted maliciously. The public ledger is offered as a
cryptographic constraint on this risk, as it would provide transparency
while also serving as a key to authorities' access
\citep[p.~556]{Green.2021ar}.

\subsubsection{Content Moderation in
E2EE}\label{content-moderation-in-e2ee}

Important efforts have gone into supporting content moderation in E2EE
messaging services. \citep{CDT.2022ol} provides an overview;
\citep{Scheffler.2023sc} supplies a systematic literature
review.\footnote{For a meticulous analysis of U.S. communications
  privacy statutory law, see \citep{Duan.2024cm}.} The present survey of
lawful-surveillance protocols does not encompass all proposals made in
this context, which, as mentioned before, is not limited to law
enforcement concerns. Some approaches, however, would be consistent with
lawful-surveillance protocol requirements to the extent that the access
to (otherwise) encrypted data would be controlled not just by policy but
also by cryptography \citep[e.g.][]{Scheffler.2023pv}.
Section~\ref{sec-apple-csam} discusses Apple's protocol, which uses
approaches that have also been explored for content moderation in E2EE
messaging services, but was instead designed at the operating system
level and for cloud-uploaded images stored on an otherwise FDE iOS
device.

\subsection{Privacy as Security and the Current State of Research on
Lawful-Surveillance Protocols}\label{sec-protocols-current-state}

Researchers with a range of views on exceptional access agree that no
proposal can currently be considered viable for implementation on the
devices of the billions of people who have smartphones
\citep{Varia.2018le, Feigenbaum:2019fy, Landau.2018bs}. This includes
proponents of lawful-surveillance protocols such as Savage, who
emphasizes that his work should be seen as a preliminary contribution
toward more advanced research \citep[p.~1772]{Savage:2018cd}. Other
proposals include the same caveat \citep[p.~562]{Green.2021ar}.

Arguing that exceptional access deserves more research, Feigenbaum
considers that ``the desire of many in computer security and related
communities for the {[}law-enforcement access{]} question to be declared
`asked and answered' and simply go away is unrealistic''
\citep[p.~29]{Feigenbaum:2019fy}. Some argue that the mere existence of
such research is problematic because it comes at the expense of research
to improve systems security, does not resolve geopolitical problems
concerning the use of these mechanisms by non-democratic states, and can
be used in the debate to make it seem that the scientific field is
divided about the need to protect strong encryption \citep{Cohn.2018ss}.
The U.S. Attorney General did, in fact, use Ozzie's proposal to defend
his position shortly thereafter in a speech at a cybersecurity
conference \citep{Barr.2019ro}. One response from researchers working on
lawful-surveillance protocol proposals is that more research is better,
enabling more grounded defenses of encryption against unacceptable
proposals \citep[p.~13]{ServanSchreiber:2020ts}.

It is beyond the scope of this work to discuss the desirability of such
research in light of the risks pointed out by its critics. I want to
draw attention to what the proponents of lawful-surveillance protocols
do \emph{not} discuss: an articulation of the right to privacy that does
not subsume it under a right to security. It is important to stress,
however, that this omission does not occur because these researchers
consider such aspects foreign to their professional responsibilities as
academics in a field not dedicated to political or philosophical
discussions. On the contrary, their work thoughtfully and explicitly
engages with such questions. And, as we have seen, the broader debate
concerning encryption and the right to privacy has, in fact, framed the
issue as a security problem, redefining the conflict as security
\emph{vs.} security, rather than privacy \emph{vs.} security.

Even the objections to the proposals examined above do not suggest
another way to view them. Thus, for example, the main controversy
regarding the proposals of Stefan Savage and Ray Ozzie---so far the most
discussed, including outside the fields of cryptography and computer
security---centers on a comparison of \emph{risks}. Savage and Ozzie
suggest that their proposals do not significantly increase the risks
inherent in operating system updates to which users are already exposed
\citep[p.~1770]{Savage:2018cd}. The objection here is that the new risks
would be of a different order, given the day-to-day operation imagined
for a lawful-surveillance system
\citep[p.~7]{Green.2018ro, Pfefferkorn:2018uf}. If, however, one limits
the system's operation to the frequency of normal operating-system
update cycles, as Varia contemplates \citep{Varia.2018le}, the objection
appears less powerful.

As for Wright and Varia's crumple zones, the objections do not point to
flaws in the design principles---described by Green, Kaptchuk, and van
Laer as ``a theoretically elegant solution''---but question whether the
proposal achieves appropriate security against malicious actors, whether
it offers transparency, and whether it would in fact satisfy
authorities' demands for access \citep[p.~555]{Green.2021ar}. The
proposal's premise---that imposing computational costs is the way to
reestablish the balance between privacy and security specified by courts
for the ``physical world'' over centuries
\citep[p.~288]{Wright:2018dj}---did not elicit objections.

This limitation of how the right to privacy is considered is found not
only in this cryptography and security scholarship but also in
public-policy analyses, such as the application of the framework
proposed in the NAS report \citep{Bellovin.2018ac}. The framework itself
sets out questions for evaluating exceptional-access schemes
\citep[pp.~87--92]{NationalAcademiesofSciencesEngineeringandMedicine:2018br}.
Concerns about security (as well as economic concerns) are clear and
translate into questions about the gains in effectiveness relative to
the costs of exceptional-access schemes. The right to privacy is
likewise presented in terms of security, in the broader sense described
above, of procedures and oversight (Section~\ref{sec-securing}). It is
broken down into sub-questions about authorization control and limiting
access, resistance to abuse and failures, and spillovers to third-party
data about people not targeted by surveillance (for example, those who
communicate with the target).

This emphasis on redefining privacy as security fails to account for the
extent to which schemes like those described above
(Section~\ref{sec-protocols-device}, Section~\ref{sec-protocols-communications})
establish structures that would have to be imposed on everyone, even
those about whom there is no suspicion whatsoever. So long as the
proposals were secure enough, the NAS framework has nothing to say. This
view does not capture encryption as ``a tool for shifting power''
\citep[p.~47]{Rogaway:2015uc}. Even so, such proposals could be
presented as meeting the specifications of privacy-as-security. A
conception of the right to privacy that takes this into account is
needed to guide the development of other protocols and systems.

\section{Why Security Does Not Capture What Is at
Stake}\label{sec-beyond-security}

While the security conception of privacy (Section~\ref{sec-securing}) is
important and compelling, it is insufficient for understanding what is
at stake. It is a dimension of the right to privacy as it pertains to
encryption regulation, but it is not the only one.

The argument, then, is not that the right to privacy should \emph{not}
also take this security into account. Yet it must stand alongside
another dimension. Understanding privacy as having more than one
dimension (or aspect) aligns with influential theories of other
fundamental rights. This is Ronald Dworkin's approach to the value of
freedom of expression, which for Dworkin has an instrumental dimension,
tied to self-government, and a constitutive dimension, tied to
democratic legitimacy
\citetext{\citealp[pp.~199--200]{Dworkin:1999wn}; \citealp[p.~356
\emph{et seq}]{Dworkin:1999uv}}.

Next, I discuss how the lawful-surveillance protocols reviewed
above---even with their challenges---raise questions that cannot be
properly answered if the right to privacy is reduced to security. I also
consider how this understanding of the right to privacy fails to explain
why government hacking would be acceptable, a stance adopted by some
critics of exceptional-access proposals. The section concludes by
considering how the response to Apple's proposed client-side scanning
system bears the seeds of a conception of privacy that goes beyond
security.

\subsection{Trading Security}\label{sec-tradingsecurity}

Discussions about exceptional access and lawful-surveillance protocols
often assess security by focusing on the changes that would ultimately
be introduced in computer systems. They take the perspective of how
systems designed according to their principles would meet officials'
demands for access under existing legal arrangements and institutional
conditions. Thus, for example, figures about encrypted devices held by
the FBI that would require unlocking are frequently cited. These numbers
were disputed: at first, the FBI spoke of ``about 7,500 mobile devices''
\citep{Rosenstein.2017ec}---a figure cited even in the NAS report
\citep[p.~41]{NationalAcademiesofSciencesEngineeringandMedicine:2018br}.
Later, it revised that number down to between one and two thousand
\citep{WaPo.2018oe}. As we have seen, the high number of devices to be
unlocked creates a complication
\citep[p.~7]{Green.2018ro, Pfefferkorn:2018uf} for the security in
managing escrowed keys in proposed protocols like Savage's
(Section~\ref{sec-savage-ozzie}). Even modifications, like Varia's
suggested mitigation (relying on normal system updates)
\citep{Varia.2018le} operate the lawful-surveillance protocol itself.

Yet encryption advocates rely not on a narrow notion of (computer)
security, but a broader notion of national or societal security
\citep{Nissenbaum.2005wc}, and rightfully so. Under this notion of
security, we could imagine hypotheticals that modify the broader
institutional context in which insecure lawful-surveillance protocols
are deployed, and yet, by offsetting other factors (outside of the
cryptographic protocols), the overall level of security is not
diminished.

For instance, U.S.~federal law allows wire interceptions (wiretaps) only
for a list of predicate offenses.\footnote{18 U.S.C. §~2516(1).} For
electronic communications, any federal felony meets the statutory
requirement.\footnote{18 U.S.C. § 2516(3).} Such statutory requirements
are often seen as ``more restrictive than what is required by the Fourth
Amendment.''\footnote{U.S. Dep't of Just., Just. Manual, §9-7.100.}
Access to the contents of a cell phone is not subject to such stricter
requirements (though it is still subject to the Fourth
Amendment\footnote{\emph{Riley v. California}, 134 S. Ct. 2473, 2493
  (2014) (``information on a cell phone is immune from search; it is
  instead that a warrant is generally required before such a search'').}).
We know that law enforcement agencies rely on tools such as those sold
by Cellebrite and Grayshift to overcome device encryption and access the
contents of cell phones \citep[p.~35]{Zinkus.2022ds}. This authority and
the more frequent use of such tools (discussed in the next subsection)
pose higher risks, resulting from the breadth of the circumstances where
surveillance is permitted.

These variables can be modified to offset the insecurity resulting from
schemes like those proposed by Savage. For instance, Title III statutory
requirements could be extended to access to data stored in personal
devices and could be subject to the limited list of predicate offenses
required for wire interception applications. This could offset the risks
presented by protocols like Savage's. In fact, the requirements could be
stricter still: access to device-stored data could be limited to Class D
felonies (punishable by more than five and up to ten years), for
example. If what we are assessing is merely the security provided by
each scenario, we must concede that changes in the legal conditions
underlying each protocol could offset the risks of adopting
it.\footnote{Hewson and Harrison caution against conflating
  institutional safeguards (represented by legal authorization
  requirements for data access) with safeguards in exceptional-access
  mechanisms, since part of the insecurity of such mechanisms is
  independent of their use \citep[p.~12]{Hewson.2022td}. This point is
  well taken and does not contradict the discussion here, which assesses
  system security in light of the risks currently posed by government
  hacking.}

Proposals like Savage's make this assessment more uncertain because the
risks they generate lie not only in insider attacks and abuses (such as
by malicious police officers and smartphone manufacturer employees) but
also in external adversaries \citep[pp.~342--343]{Corn:2017jh}. Wright
and Varia's proposal (Section~\ref{sec-crumple-zones}) is perhaps the
most plausible for gauging such offsetting of insecurity, because the
high entry cost of the first crumple zone (hundreds of millions to
billions of dollars) would exclude a significant share of external
adversaries. Thus, changes in the legal requirements for obtaining
stored data could more clearly compensate for insecurity added by
mechanisms like Wright and Varia's.

These considerations are only offered here to show that there must be
more to our understanding of the value of privacy as it implicates
encryption. I am not endorsing the adoption of any lawful-surveillance
protocol; the statutory changes are hypothetical. The point is to show
that a securitized right to privacy would say nothing against adopting
one of them in these scenarios.

The last section shows that the truce in the Crypto Wars should be seen
as tenuous. It would last only as long as there is no proposal strong
enough to convince the community (or part of it) of its security. This
subsection has sought to show that opponents' position on
exceptional-access mechanisms is precarious even today---at least if it
is to be supported solely by privacy-as-security.

\subsection{Government Hacking: Tolerated
Insecurity}\label{sec-govthacking}

Privacy-as-security is also not enough to explain the stance of
opponents of exceptional access on another issue, government
hacking.\footnote{Government hacking is discussed here insofar as it is
  relevant to the argument. For an overview of the current state of
  affairs in government hacking and the regulation of zero-day
  vulnerabilities, see
  \citep{Mayer:2018vc, Haber.2024lt, Fidler.2024zp}.}

Government hacking has been offered as a way out of the debates over
exceptional access
\citetext{\citealp[p.~328]{LiguoriFilho:2020vq}; \citealp[pp.~1259--1260]{Bercovitz.2021le}}.
It is described as a way for the ``government to continue to support
strengthening encryption and simultaneously give law enforcement
resources to bolster their capabilities to conduct investigations in an
environment of evolving technology and strong encryption''
\citep[p.~12]{Finklea.2017lh}. Its adoption is even seen as a direct
consequence of the absence of exceptional-access solutions
\citep[p.~127]{Mayer:2018vc}. Even critics of exceptional-access schemes
and of imposing assistance duties on developers and manufacturers of
products that employ encryption accept the use of government hacking
\citetext{\citealp[p.~66]{Nguyen:2017wp}; \citealp[\emph{but
see}][]{Pfefferkorn:2018wx}}, which they regard as a preferable
alternative. An article authored by several leading critics of
exceptional-access mechanisms frames the issue as follows
\citep[p.~5]{Bellovin:I2sir3KP}:

\begin{quote}
Put simply, the choice is between formalizing (and thereby constraining)
the ability of law enforcement to occasionally use existing security
vulnerabilities---something the FBI and other law enforcement agencies
already do when necessary without much public or legal scrutiny--- or
living with those vulnerabilities and intentionally and systematically
creating a set of predictable new vulnerabilities that despite best
efforts will be exploitable by everyone.
\end{quote}

To discipline and limit the risks of government hacking, the authors
argue that the tools should be regulated to control their proliferation,
that vulnerabilities should generally be disclosed to developers or
manufacturers so they can be patched, and that there should be
legislative, judicial, and public oversight. Thus, given that
vulnerabilities will always exist \citep[pp.~27--30]{Bellovin:I2sir3KP},
they argue that the risks associated with government hacking are
acceptable.

However, the existence of hacking tools does not affect only the
security of investigation targets, as the authors observe \citep[see
also][pp.~66--67]{Nguyen:2017wp}. Even with strict controls in place,
there will always be a risk that such tools will be diverted by insiders
or appropriated by external adversaries \citep[p.~55067]{Li:2018kp}.
This possibility poses a security risk for everyone who uses the device
or system for which the hacking tool was designed. That risk is not the
same as the risk already present in the vulnerabilities that enable such
tools. The same reasoning we saw above applies here with respect to the
difference between risks in keys for signing operating-system updates
and unlock keys like those envisioned in the proposals by Savage and
Ozzie. Vulnerabilities---even publicly known ones
\citep[p.~899]{Koops:2018bh}---still depend on the development of an
exploit so that an attack can be launched (for example, to extract data
sought by the police). Government hacking tools, whether developed by
its agencies or acquired from third parties, provide that utility.

These tools may not have the same immediate utility as an unlock key,
particularly in the case of exploits, which would still require
knowledge likely beyond the reach of ordinary criminals. But recall that
schemes like those of Savage and Ozzie would likewise not grant
immediate access to encrypted data---the device would still have to be
physically seized. If the police uses commercially developed tools,
these risks become even greater. And police do, in fact, use commercial
products \citep[p.~607]{Uresk.2020cd}. And the risks are not
hypothetical. In 2017, a breach allowed the extraction of 900GB from
Cellebrite's servers, including access credentials and, reportedly, also
evidence \citep{Cox.2017ch}. The source code of another tool, developed
by a Cellebrite competitor, Grayshift, was allegedly obtained by actors
who then extorted the company into not publishing it
\citep{Vice.2018gs}. A report on the use of such tools in the United
States concluded that more than 2,000 agencies acquired one, including
the fifty largest local police forces and all state forces
\citetext{\citealp{Upturn.2020me}; \citealp[see
also][]{Quinlan:2016wr}}.

Bellovin, Blaze, Clark, and Landau also offer another reason to think
government hacking poses less risk than the deliberate introduction of
vulnerabilities. Even if vulnerabilities expose any given target, it
would be much harder to conduct operations against all members of a
large population, they argue \citep[p.~64]{Bellovin:I2sir3KP}. Once
again, we should note that this would not necessarily apply to
self-escrow proposals like those of Savage and Ozzie, which involve the
``deliberate introduction of vulnerabilities'' (the exceptional-access
mechanism via self-escrowed keys) but mitigate the risk of abuse for
mass surveillance by requiring physical seizure of the device in order
to access encrypted data.

That a securitized right to privacy is not sufficient to address the
problem becomes even clearer when we consider variations in the legal
conditions for access to encrypted data while weighing the impact of
government hacking in fostering a market for hacking tools. Upturn's
report on the use of tools such as those sold by Cellebrite and
Grayshift estimates millions of dollars in spending on products from
these two firms and other vendors \citep{Upturn.2020me}. The purchase of
such tools with public funds is an important source of revenue for this
market \citep[ p.~8]{Hewson.2022td}.

From the standpoint of security alone, a scenario in which the police
and other authorities stop buying these products while a self-escrow
obligation is adopted could counterintuitively yield a \emph{positive}
balance for privacy. Consider that withholding current public spending
on hacking tools could weaken the market that now incentivizes keeping
vulnerabilities in important systems (such as mobile operating systems)
secret from the public and from the systems' developers. Without being
patched by developers, these vulnerabilities pose risks and reduce user
security. The economic cooling of the hacking tool market (resulting
from the elimination of a major revenue source) could then lead to
greater security, especially against malicious third parties (though not
as much against insiders, such as rogue police officers). If it is true
that a self-escrow obligation (as in Savage's and Ozzie's approach)
would represent less security, the balance between the increase
resulting from ending purchases of hacking tools and the decrease
resulting from self-escrow could be positive. In other words, taken
together, these two measures could offset one another---or even reduce
the insecurity to which users are currently subject.

Thus, if the right to privacy is reduced to risks and security, a law
that created a self-escrow obligation while prohibiting police
acquisition of hacking tools would not merely avoid violating the right
to privacy---it would be described as an expansion of that right. The
result of reasoning limited to risks and security is that this radical
transformation in state surveillance capacities would not even raise
problems in terms of the right to privacy.

One objection to the argument here would be to question whether these
two measures together (a self-escrow obligation and a ban on the state's
acquisition of hacking tools) would, in fact, yield a positive (or
neutral) security balance. The argument does not depend on this,
however. It is possible that the balance would be negative. In fact,
designing a method for rigorously assessing this balance would be
difficult \citep{Pozen.2016pp}, perhaps even unfeasible, both because of
uncertainties regarding information security and the difficulty of
establishing an objective method (not biased toward a given outcome) for
calculating the factors at play. My goal is not to design or test any
such method. What I want to stress is that it would be a mistake to
think that the impact on the right to privacy would be decided solely on
the basis of this calculation.

\subsection{Apple's NeuralHash and Expanded Protections for
Children}\label{sec-apple-csam}

In August 2021, the same Apple that had confronted the FBI made an
announcement that was seen as the antithesis of its stance in the San
Bernardino case \citep{Landau.2022ns}. Among other measures for
``expanded protections for children'' in the next versions of its mobile
and desktop operating systems,\footnote{The first involved using machine
  learning to analyze photos that children using iMessage attempted to
  send, in order to prevent them from sharing ``sexually explicit
  content'' \citep{Weigel.2022cs}. The second enhanced the Siri virtual
  assistant to provide guidance on child sexual abuse material (CSAM).
  The third, a CSAM detection system, is discussed here.} ~the company
introduced a tool to scan photos for child sexual abuse material (CSAM)
\citep{Apple.2021pc}. Of course, many other providers already deploy
solutions to detect CSAM on their services \citep{Wired.2021at};
PhotoDNA, developed by Microsoft, is one of the most widespread
\citetext{\citealp[p.~136]{Common.2020fr}; \citealp[
p.~12]{Farid.2022ph}}. The difference in Apple's case lay in how this
mechanism would work.

Whereas other tools operated on servers scanning content stored in the
cloud, Apple's system would run directly on the device itself
\citep{Stamos.2021ap}. It would apply only to photos stored in iCloud,
but it would be client-side. In addition to the on-device system, a
second layer of verification would take place server-side, on Apple's
servers, eliminating false positives and limiting attacks compared to a
model in which the entire system operated only on the device
\citep[p.~37]{Abelson.2022bp}. A manual review would follow as a third
layer. These latter two layers would only be triggered once the
on-device system had identified a certain minimum number of images
\citep[p.~4]{Apple.2021ts}. That threshold was initially set at 30
images \citep[p.~36]{Abelson.2022bp}, though Apple retained the ability
to modify it \citep[p.~37]{Abelson.2022bp}.

The move came after Apple had been publicly criticized as one of the
companies that submitted the fewest CSAM reports to authorities
\citep{NYT.2019ca, NYT.2020cs}. Given this, and considering it was
reported that the company had previously been pressured by the FBI not
to adopt end-to-end encryption for iCloud \citep{Reuters.2020af}, the
speculation was that the CSAM detection system was a way for Apple to
allay concerns in preparation for rolling out an encrypted iCloud. This
hypothesis is strengthened by the fact that nothing prevented Apple from
scanning files server-side \citep[ (quoting Matt Green)]{Wired.2021at}.

What is striking in this scenario is that Apple faced a choice between
maintaining the status quo, in which all user files in iCloud could be
accessed from the server---and were therefore available both to
officials and to malicious actors---or moving to a scenario where the
files would be encrypted under a scheme that could be seen as a form of
lawful-surveillance protocol. Apple argued that its client-side scanning
system was superior to existing server-side alternatives, which
``create{[}{]} privacy risk for all users'' \citep[ p.~4]{Apple.2021fa}.
It boasted that its system ``provide{[}d{]} significant privacy benefits
over those techniques by preventing Apple from learning anything about
photos unless they both match to known CSAM images and are included in
an iCloud Photos account that contains a collection of known CSAM''
\citep[p.~4]{Apple.2021fa}. To back up its claims, Apple published
statements from three respected cryptographers, each offering positive
evaluations of the system
\citep{Pinkas.2021ap, Forsyth.2021ap, Bellare.2021ap}. David Forsyth
concluded that ``Apple's approach preserves privacy better than any
other I am aware of'' \citep[p.~2]{Forsyth.2021ap}. Mihir Bellare
described the wholesale scanning of all photos (the CSAM detection
mechanism against which Apple's model was contrasted) as a means of
restricting CSAM that fails to respect the fact that ``our photos are
personal, recording events, moments, and people in our lives.'' He
argued that Apple had ``found a way to detect and report CSAM offenders
while still respecting these privacy constraints''
\citep[p.~1]{Bellare.2021ap}.

Even the \emph{Bugs in our pockets} report \citep{Abelson.2022bp}
(authored by some of the figures behind the influential \emph{Risks}
\citep{Abelson.1997} and \emph{Keys under doormats}
\citep{Abelson:2015bl} reports) does not go so far as to say that a
system like Apple's would be riskier than the alternative of server-side
scanning. That does not mean the system had no known vulnerabilities
\citep{Sparkes.2021pf}: the report noted
\citep[pp.~12--13]{Abelson.2024bo} that within 48 hours of NeuralHash's
code being published, researchers had already been able to generate
artificially colliding images to create false positives and to
manipulate images with imperceptible perturbations to significantly
alter the perceptual hash, leading to false negatives. Apple responded
by stating it was prepared for collisions and that images that met the
threshold of the client-side-based system would go through a secondary,
undisclosed classifier before proceeding to manual review
\citep{Brandom.2021as}.

The report raised numerous concerns about the system's
effectiveness---which could even be undermined if its vulnerabilities
were manipulated to flood it---and about its potential abuse or
expansion beyond CSAM \citep{Strupekk.2022ph}. But it did not claim that
the proposed Apple system was less secure or privacy-protecting than the
server-side approach.\footnote{This comparison would be cumbersome. The
  assessments that Apple published with its announcement
  \citep{Pinkas.2021ap, Bellare.2021ap, Forsyth.2021ap} analyze privacy
  protections compared to a traditional server-side mechanism and with
  regard to targeted data. Yet the \emph{Bugs} report notes \citep[
  pp.~10--11]{Abelson.2024bo} that the large-scale introduction of a
  client-side scanner creates complexity to OS security and creates a
  new attack surface. A proper comparison would have to account for this
  source of insecurity, thereby adding a second dimension to the
  security calculus. That is not the goal of this article.}

Its most fundamental concern seemed to lie elsewhere. After noting that
the economic dimension of the system also had to be considered (because
Apple's tool would make surveillance cheap, while democratic societies
make searches costly), the report posed the following question \citep[
p.~16]{Abelson.2024bo}:

\begin{quote}
The proposal to preemptively scan all user devices for targeted content
is far more insidious than earlier proposals for key escrow and
exceptional access. Instead of having targeted capabilities such as to
wiretap communications with a warrant and to perform forensics on seized
devices, the agencies' direction of travel is the bulk scanning of
everyone's private data, all the time, without warrant or suspicion.
That crosses a red line. Is it prudent to deploy extremely powerful
surveillance technology that could easily be extended to undermine basic
freedoms?
\end{quote}

If we assume the answer might be No, we must acknowledge that this
discussion cannot take place if the right to privacy is reduced to
security. That reading is precisely what underpins the
privacy-protective assessments of Bellare and Forsyth. Although this
framing has dominated recent debates on encryption regulation, it is
limited, as shown above. The authors touch on a range of important
issues that escape security. I am inclined to agree, for instance, that
it matters, as suggested by the report's title, \emph{Bugs in our
pockets}, that the scanning takes place on a device so central to our
private lives. There seems to be something intrusive about the user
having a system that treats them as an adversary running on their phone.
Given that the targeted content is contraband over which no privacy
interest may be asserted and how narrow the category of targeted content
is (CSAM), any legitimate privacy concern would have to be located at a
meta-level, flowing from a normative ascription of privacy to the
medium. This would be similar to the approach communications privacy
protections take, yet devices are presumably not covered by
communications privacy provisions, and any extension may raise
complications for ordinary computer searches. The system's
population-wide reach and its constant, suspicionless operation likewise
strike me as compelling. I mention these directions not to pursue them
here, but again to underscore that they require developing a conception
of privacy relevant to encryption policy that accounts for more than
security.

\section*{Conclusion}\label{sec-conclusion}
\addcontentsline{toc}{section}{Conclusion}

Encryption policy is certainly better off when the right to privacy is
not unjustifiably assumed to impose costs on security. Commentators who
argued that weakening encryption can also impose costs on security have
made an important contribution to public debate. Much of the credit is
owed to cryptography and security researchers. And there is, of course,
good strategy in this move, which speaks about privacy in a language
that accepts the normative priorities held by government officials
seeking exceptional access. Security is not just more strategic because
of that; it also seems more quantifiable and, hence, objective.

Yet the security conception of privacy that has driven much of the
debate is limited. Success in lawful-surveillance protocols research,
broadly consistent with this conception of privacy, would push it to a
breaking point. Privacy as security is challenged not just by this
prospect, however uncertain it might be. Even the current landscape
cannot be fully explained by security alone. The very response to
Apple's now-abandoned client-side scanning proposal intimates as much.

The path ahead as such requires understanding privacy beyond security.
This does not mean abandoning security. Any good conceptual account will
have to include security as a dimension of the value of privacy. In
fact, it is possible that a better conception of privacy will also
entail a more normatively grounded, context-sensitive\footnote{As
  Scheffler and Mayer \citeyearpar[ p.~426]{Scheffler.2023sc} rightfully
  emphasize in the context of content moderation in E2EE messaging
  services, seizing on opinion survey responses about TLS inspection,
  which showed more support to TLS proxies used by elementary schools
  and employers \citep{ONeill.2017ti}.} conception of security
itself.\footnote{As Nissenbaum \citeyearpar[p.~64]{Nissenbaum.2005wc}
  observes, ``{[}T{]}he quest for computer security has moral force only
  to the extent that it promotes the common value of freedom from harm.
  In other words, the issue is not merely why these are classifiable as
  security concerns but why people \emph{deserve}, or have a right, to
  be thus secured.''} There is no reason to believe that this exercise
should end back where we started, pitting privacy and security against
each other
\citetext{\citealp[pp.~159--162]{Dworkin:2006wo}; \citealp{Dworkin:2011tm}}.

I have not put forward any such conception here. My goal has been to
show that there is need for it. A richer conceptual understanding is
required not just to assess policy and regulation. Work on
lawful-surveillance protocols will also benefit from this; alternative
conceptions might help revise the specifications for this agenda. It
goes without saying that this project has a better outlook with
cross-cutting collaboration, as no useful conception of privacy will
ignore what is technologically or mathematically viable, and useful
protocols are more likely to yield from well-informed, richer
conceptions of privacy. Though normative insight and understanding are
not bound by academic disciplines, cryptography and security researchers
have pushed forward largely unaided. Legal scholars and other privacy
theorists seem to have much work to do.

\begin{acks}

I thank the anonymous reviewers. I am deeply grateful to Susan Landau
for her exacting reading and criticism of an earlier draft, which I
sought to clarify in this final version. My thanks to the participants
of the 2025 Fellows Conference at Yale Law School's Information Society
Project.

This article revises and updates parts of my PhD dissertation
\citep{APLM.2022pc}. I thank the committee (Diego F. Aranha, Laura
Schertel Mendes, Marcel Leonardi, Marta Saad, and Rafael Mafei), and my
supervisor, Virgílio Afonso da Silva, to whom I am indebted.

\end{acks}

\bibliography{refs.bib}

@article{Abelson:2015bl,
  title = {Keys under Doormats: Mandating Insecurity by Requiring Government Access to All Data and Communications},
  year = 2015,
  month = nov,
  journal = {Journal of Cybersecurity},
  volume = {1},
  number = {1},
  pages = {1--11},
  doi = {10.1093/cybsec/tyv009},
  author = {Abelson, Harold and Anderson, Ross and Bellovin, Steven M. and Benaloh, Josh and Blaze, Matt and Diffie, Whitfield and Gilmore, John and Green, Matthew and Landau, Susan and Neumann, Peter G. and Rivest, Ronald L. and Schiller, Jeffrey I. and Schneier, Bruce and Specter, Michael A. and Weitzner, Daniel J.}
}

@article{Abelson.1997,
  title = {The Risks of Key Recovery, Key Escrow, and Trusted Third-Party Encryption},
  year = 1997,
  month = may,
  journal = {World Wide Web Journal},
  volume = {2},
  number = {3},
  pages = {241--257},
  
  author = {Abelson, Hal and {Ross Anderson} and M Bellovin, Steven and Benaloh, Josh and Blaze, Matt and {Whitfield Diffie} and {John Gilmore} and G Neumann, Peter and L Rivest, Ronald and Schiller, Jeffrey I. and Schneier, Bruce}
}

@misc{Abelson.2022bp,
  title = {Bugs in Our Pockets: The Risks of Client-Side Scanning},
  year = 2021,
  number = {2110.07450},
  eprint = {2110.07450},
  publisher = {arXiv},
  archiveprefix = {arXiv},
  
  author = {Abelson, Hal and Anderson, Ross and Bellovin, Steven M. and Benaloh, Josh and Blaze, Matt and Callas, Jon and Diffie, Whitfield and Landau, Susan and Neumann, Peter G. and Rivest, Ronald L. and Schiller, Jeffrey I. and Schneier, Bruce and Teague, Vanessa and Troncoso, Carmela}
}

@article{Abelson.2024bo,
  title = {Bugs in Our Pockets: The Risks of Client-Side Scanning},
  shorttitle = {Bugs in Our Pockets},
  year = 2024,
  journal = {Journal of Cyber Security},
  volume = {10},
  number = {1},
  doi = {10.1093/cybsec/tyad020},
  urldate = {2025-09-30},
  author = {Abelson, Harold and Anderson, Ross and Bellovin, Steven M. and Benaloh, Josh and Blaze, Matt and Callas, Jon and Diffie, Whitfield and Landau, Susan and Neumann, Peter G. and Rivest, Ronald L. and Schiller, Jeffrey I. and Schneier, Bruce and Teague, Vanessa and Troncoso, Carmela},
  month = {January 1,}
}

@article{Abreu:2018ga,
  title = {Disrupting the Disruptive: Making Sense of App Blocking in {{Brazil}}},
  year = 2018,
  month = jul,
  journal = {Internet Policy Review},
  volume = {7},
  number = {3},
  doi = {10.14763/2018.3.928},
  author = {Abreu, Jacqueline de Souza}
}

@article{Ahmad:2009ce,
  title = {Restrictions on Cryptography in {{India}} -- {{A}} Case Study of Encryption and Privacy},
  year = 2009,
  journal = {Computer Law \& Security Review},
  volume = {25},
  number = {2},
  pages = {173--180},
  doi = {10.1016/j.clsr.2009.02.001},
  author = {Ahmad, Nehaluddin}
}

@misc{AMS.2023la,
  title = {Landau Awarded 2024 {{Bertrand Russell Prize}}},
  year = 2023,
  journal = {American Mathematical Society},
  author = {{American Mathematical Society}},
  month = {November 7,},
  note = {Retrieved from https://www.ams.org/news?news\_id=7241},
  howpublished = {American Mathematical Society}
}

@book{Anderson.2020,
  title = {Security Engineering: A Guide to Building Dependable Distributed Systems},
  year = 2020,
  edition = {3},
  publisher = {Wiley},
  address = {Indianapolis, IN},
  
  author = {Anderson, Ross}
}

@article{Anderson.2021re,
  type = {Review Of},
  title = {Crypto Wars --- the Fight for Privacy in the Digital Age: A Political History of Digital Encryption},
  year = 2023,
  journal = {Cryptologia},
  volume = {47},
  number = {3},
  pages = {285--298},
  doi = {10.1080/01611194.2021.2002977},
  author = {Anderson, Patrick D.},
  collaborator = {{Craig Jarvis}}
}

@incollection{APLM.2023pc,
  title = {Privacy at a Crossroads},
  booktitle = {Research Handbook on Law and Technology},
  year = 2023,
  pages = {214--221},
  publisher = {Elgar},
  address = {Cheltenham},
  author = {Monteiro, Artur Pericles Lima},
  editor = {Bartosz, Bro{\.z}ek and Kanevskaia, Olia and Przemys{\l}aw, Pa{\l}ka}
}

@misc{Apple.2021fa,
  title = {Expanded Protections for Children: Frequently Asked Questions},
  year = 2021,
  urldate = {2022-05-10},
  author = {{Apple}},
  note = {Retrieved from https://web.archive.org/web/20210809022217/https://www.apple.com/child-safety/pdf/Expanded\_Protections\_for\_Children\_Frequently\_Asked\_Questions.pdf}
}

@misc{Apple.2021pc,
  title = {Expanded Protections for Children},
  year = 2022,
  author = {{Apple}},
  month = {August 28,},
  note = {Retrieved from https://web.archive.org/web/20210828174414/https://www.apple.com/child-safety/}
}

@misc{Apple.2021ts,
  title = {{{CSAM}} Detection: Technical Summary},
  year = 2021,
  urldate = {2022-05-28},
  author = {{Apple}},
  note = {Retrieved from https://www.apple.com/child-safety/pdf/CSAM\_Detection\_Technical\_Summary.pdf}
}

@misc{Ashbrook.2016mh,
  title = {Michael {{Hayden}}: {{America}} Is Safer with End-to-End Encryption},
  year = 2016,
  month = {March 1,},
  note = {Retrieved from https://www.wbur.org/onpoint/2016/03/01/michael-hayden-nsa-encryption},
  author = {Ashbrook, Tom},
  howpublished = {On Point, WBUR}
}

@misc{Backer.2022ms,
  title = {My Spy Is Always with Me},
  year = 2022,
  journal = {Verfassungsblog},
  doi = {10.17176/20220818-181949-0},
  urldate = {2026-01-16},
  copyright = {Creative Commons Attribution Share Alike 4.0 International},
  author = {B{\"a}cker, Matthias and Buermeyer, Ulf},
  month = {August 18,},
  note = {Retrieved from https://verfassungsblog.de/my-spy-is-always-with-me/},
  howpublished = {Verfassungsblog}
}

@article{Baker.1994dw,
  title = {Don't Worry Be Happy},
  year = 1994,
  journal = {Wired},
  urldate = {2026-01-30},
  chapter = {tags},
  author = {Baker, Stewart A.},
  month = {June 1,},
  note = {Retrieved from https://www.wired.com/1994/06/nsa-clipper/}
}

@book{Bamford.1983pp,
  title = {The Puzzle Palace},
  year = 1983,
  publisher = {Penguin Books},
  address = {Boston},
  
  author = {Bamford, James}
}

@article{Banisar:1999tr,
  title = {Stopping Science: The Case of Cryptography},
  year = 1999,
  journal = {Health Matrix},
  volume = {9},
  number = {2},
  pages = {253--287},
  author = {Banisar, David}
}

@misc{Barr.2019ro,
  title = {Attorney {{General William P}}. {{Barr}} Delivers Keynote Address at the {{International Conference}} on {{Cyber Security}}},
  year = 2019,
  publisher = {U.S. Department of Justice},
  howpublished = {https://www.justice.gov/opa/speech/attorney-general-william-p-barr-delivers-keynote-address-international-conference-cyber},
  author = {{U.S. Department of Justice}},
  month = {July 23,},
  note = {Retrieved from https://www.justice.gov/opa/speech/attorney-general-william-p-barr-delivers-keynote-address-international-conference-cyber}
}

@article{Basri.2025pr,
  title = {{\emph{Podchasov v }}{{{\emph{Russia}}}} : A New Frontier in the Crypto-Wars before the {{Strasbourg Court}}},
  shorttitle = {{\emph{Podchasov v }}{{{\emph{Russia}}}}},
  year = 2025,
  journal = {International Data Privacy Law},
  doi = {10.1093/idpl/ipaf031},
  urldate = {2026-01-18},
  copyright = {https://creativecommons.org/licenses/by/4.0/},
  author = {Basri, Nahide},
  month = {December 5,}
}

@misc{Bellare.2021ap,
  title = {The {{Apple PSI}} Protocol},
  year = 2021,
  urldate = {2022-05-10},
  author = {Bellare, Mihir},
  month = {July 30,},
  note = {Retrieved from https://web.archive.org/web/20210805192048/https://www.apple.com/child-safety/pdf/Technical\_Assessment\_of\_CSAM\_Detection\_Mihir\_Bellare.pdf}
}

@article{Bellovin:I2sir3KP,
  title = {Lawful Hacking: Using Existing Vulnerabilities for Wiretapping on the Internet},
  year = 2014,
  journal = {Northwestern Journal of Technology and Intellectual Property},
  volume = {12},
  number = {1},
  pages = {1--64},
  author = {Bellovin, Steven M. and Blaze, Matt and Clark, Sandy and Landau, Susan}
}

@misc{Bellovin.2006si,
  title = {Security Implications of Applying the Communications Assistance to Law Enforcement Act to Voice over {{IP}}},
  year = 2006,
  howpublished = {Information Technology Association of America},
  note = {Retrieved from https://doi.org/10.7916/D8VT1ZV7},
  author = {Bellovin, Steven and Blaze, Matt and Brickell, Ernest and Brooks, Clinton and Cerf, Vinton and Diffie, Whitfield and Microsystems, Sun and Landau, Susan and Microsystems, Sun and Peterson, Jon and Treichler, John},
  month = {June 13,},

}

@techreport{Bellovin.2018ac,
  title = {Analysis of the {{CLEAR}} Protocol per the {{National Academies}}' Framework},
  year = 2018,
  month = may,
  number = {CUCS-003-18},
  address = {New York},
  institution = {Columbia University, Department of Computer Science},
  
  author = {Bellovin, Steven M. and Blaze, Matt and Boneh, Dan and Landau, Susan and Rivest, Ronald L.},
  type = {Report}
}

@misc{Bellovin.2018sf,
  title = {Ray {{Ozzie}}'s Proposal: Not a Step Forward},
  year = 2018,
  journal = {SMBlog},
  author = {Bellovin, Steven M.},
  month = {April 25,},
  note = {Retrieved from https://www.cs.columbia.edu/\textasciitilde smb/blog/2018-04/2018-04-25.html},
  howpublished = {SMBlog}
}

@article{Bellovin.2018tr,
  title = {Ray {{Ozzie}}'s Crypto Proposal---a Dose of Technical Reality},
  year = 2018,
  journal = {Ars Technica},
  author = {Bellovin, Steven M. and Blaze, Matt and Boneh, Dan and Landau, Susan and Rivest, Ronald L.},
  month = {May 7,},
  note = {Retrieved from https://arstechnica.com/information-technology/2018/05/op-ed-ray-ozzies-crypto-proposal-a-dose-of-technical-reality/}
}

@article{Bercovitz.2021le,
  title = {Law Enforcement Hacking},
  year = 2021,
  journal = {Columbia Law Review},
  volume = {121},
  number = {4},
  pages = {1251--1288},
  
  author = {Bercovitz, Rachel}
}

@incollection{Bernstein.2016ec,
  title = {Dual {{EC}}: A Standardized Back Door},
  booktitle = {The New Codebreakers: Essays Dedicated to {{David Kahn}} on the Occasion of His 85th Birthday},
  year = 2016,
  series = {Lecture Notes in Computer Science},
  number = {9100},
  pages = {256--281},
  publisher = {Springer},
  address = {Berlin},
  
  author = {Bernstein, Daniel J. and Lange, Tanja and Niederhagen, Ruben},
  editor = {Ryan, Peter Y. A. and Naccache, David and Quisquater, Jean-Jacques}
}

@article{Blaze.1994pf,
  title = {Protocol Failure in the Escrowed Encryption Standard},
  year = 1994,
  journal = {Proceedings of the 2nd ACM Conference on Computer and communications security - CCS '94},
  pages = {59--67},
  doi = {10.1145/191177.191193},
  
  author = {Blaze, Matt}
}

@inproceedings{Blaze.2011ke,
  title = {Key Escrow from a Safe Distance: Looking Back at the {{Clipper Chip}}},
  shorttitle = {Key Escrow from a Safe Distance},
  booktitle = {Proceedings of the 27th {{Annual Computer Security Applications Conference}}},
  year = 2011,
  pages = {317--321},
  publisher = {ACM},
  address = {New York},
  doi = {10.1145/2076732.2076777},
  urldate = {2026-01-26},
  
  author = {Blaze, Matt},
  month = {December 5,},
  note = {Retrieved from https://dl.acm.org/doi/10.1145/2076732.2076777}
}

@article{Brandom.2021as,
  title = {Apple Says Collision in Child-Abuse Hashing System Is Not a Concern},
  year = 2021,
  journal = {The Verge},
  urldate = {2026-01-30},
  author = {Brandom, Russell},
  month = {August 18,},
  note = {Retrieved from https://www.theverge.com/2021/8/18/22630439/apple-csam-neuralhash-collision-vulnerability-flaw-cryptography}
}

@misc{Burman.2021ei,
  title = {Understanding the Encryption Debate in {{India}}},
  year = 2021,
  howpublished = {Carnegie Endowment for International Peace},
  
  author = {Burman, Anirudh and Jha, Prateek},
  month = {September 13,},
  note = {Retrieved from https://carnegieendowment.org/research/2021/09/understanding-the-encryption-debate-in-india},
  
}

@misc{Caproni.2011gd,
  title = {Going Dark: Lawful Electronic Surveillance in the Face of New Technology},
  year = 2011,
  number = {112--59},
  author = {Caproni, Valerie},
  month = {February 17,},
  howpublished = {Hearing before the Subcomm. on Crime, Terrorism, and Homeland Security of the H. Comm. on the Judiciary, 112th Cong.}
}

@misc{CDT.2022ol,
  title = {Outside Looking in: Approaches to Content Moderation in End-to-End Encrypted Systems},
  shorttitle = {Outside Looking In},
  year = 2020,
  howpublished = {Center for Democracy \& Technology},
  
  author = {Kamara, Seny and Knodel, Mallory and Llans{\'o}, Emma and Nojeim, Greg and Qin, Lucy and Thakur, Dhanaraj and Vogus, Caitlin},
  note = {Retrieved from https://cdt.org/wp-content/uploads/2021/08/CDT-Outside-Looking-In-Approaches-to-Content-Moderation-in-End-to-End-Encrypted-Systems-updated-20220113.pdf},
  
}

@incollection{Charlesworth.2007pe,
  title = {Munitions, Wiretaps and {{MP3s}}: The Changing Interface between Privacy and Encryption Policy in the Information Society},
  shorttitle = {Munitions, Wiretaps and {{MP3s}}},
  booktitle = {The History of Information Security: A Comprehensive Handbook},
  year = 2007,
  pages = {771--817},
  publisher = {Elsevier},
  address = {Amsterdam \& London},
  doi = {10.1016/B978-044451608-4/50029-8},
  author = {Charlesworth, Andrew},
  editor = {de Leeuw, Karl and Bergstra, Jan}
}

@article{Clark.2025er,
  title = {{{EU}} Races to Pass New Law to Combat Online Child Abuse},
  year = 2025,
  journal = {Politico},
  urldate = {2026-01-29},
  author = {Clark, Sam},
  month = {November 26,},
  note = {Retrieved from https://www.politico.eu/article/eu-speed-up-to-pass-chat-control-bill-online-child-sexual-abuse/}
}

@misc{Cohn.2018ss,
  title = {Resisting Law Enforcement's Siren Song: A Call for Cryptographers to Improve Trust and Security},
  year = 2018,
  journal = {Lawfare},
  author = {Cohn, Cindy},
  month = {November 30,},
  note = {Retrieved from https://www.lawfaremedia.org/article/resisting-law-enforcements-siren-song-call-cryptographers-improve-trust-and-security},
  howpublished = {Lawfare}
}

@misc{Comey.2016ep,
  type = {Speech},
  title = {Expectations of Privacy: Balancing Liberty, Security, and Public Safety},
  shorttitle = {Expectations of Privacy},
  year = 2016,
  address = {Gambier, OH},
  urldate = {2026-01-30},
  author = {Comey, James},
  month = {April 6,},
  note = {Retrieved from https://www.fbi.gov/news/speeches/expectations-of-privacy-balancing-liberty-security-and-public-safety}
}

@article{Common.2020fr,
  title = {Fear the {{Reaper}}: How Content Moderation Rules Are Enforced on Social Media},
  year = 2020,
  journal = {International Review of Law, Computers \& Technology},
  volume = {34},
  number = {2},
  pages = {1--27},
  doi = {10.1080/13600869.2020.1733762},
  
  author = {Common, MacKenzie F.}
}

@misc{Cook.2016mo,
  title = {A Message to Our Customers},
  year = 2016,
  journal = {Apple},
  urldate = {2026-01-29},
  howpublished = {http://www.apple.com/customer-letter/},
  author = {Cook, Tim},
  month = {February 16,},
  note = {Retrieved from http://www.apple.com/customer-letter/}
}

@incollection{Corn:2017jh,
  title = {``{{Going}} Dark'': Encryption, Privacy, Liberty, and Security in the ``Golden Age of Surveillance''},
  booktitle = {The {{Cambridge}} Handbook of Surveillance Law},
  year = 2017,
  pages = {330--371},
  publisher = {Cambridge University Press},
  address = {Cambridge},
  doi = {10.1017/9781316481127.015},
  author = {Corn, Geoffrey S. and {Brenner-Beck}, Dru},
  editor = {Gray, David and Henderson, Stephen E.}
}

@article{Cox.2017ch,
  title = {Hacker Steals {{900GB}} of {{Cellebrite}} Data},
  year = 2017,
  journal = {Vice},
  author = {Cox, Joseph},
  month = {January 12,},
  note = {Retrieved from https://www.vice.com/en/article/3daywj/hacker-steals-900-gb-of-cellebrite-data}
}

@misc{CPSR.1994cc,
  title = {Electronic Petition to Oppose {{Clipper}}},
  year = 1994,
  urldate = {2022-05-10},
  author = {{Computer Professionals for Social Responsibility}},
  month = {January 24,},
  note = {Retrieved from https://archive.epic.org/crypto/clipper/cpsr\_electronic\_petition.html}
}

@article{Dakoff:1996wz,
  type = {Note},
  title = {The {{Clipper Chip}} Proposal: Deciphering the Unfounded Fears That Are Wrongfully Derailing Its Implementation},
  year = 1996,
  journal = {John Marshall Law Review},
  volume = {29},
  pages = {475--498},
  
  author = {Dakoff, Howard S.}
}

@article{Davis.2022da,
  title = {Decrypting {{Australia}}'s `{{Anti-Encryption}}' Legislation: {{The}} Meaning and Effect of the `Systemic Weakness' Limitation},
  year = 2022,
  journal = {Computer Law \& Security Review},
  volume = {44},
  doi = {10.1016/j.clsr.2022.105659},
  author = {Davis, Peter Alexander Earls}
}

@article{Davis.2024re,
  title = {A Right to Encryption in the {{European Union}}'s {{Charter}} of {{Fundamental Rights}}},
  year = 2024,
  journal = {Columbia Journal of European Law},
  volume = {30},
  number = {1},
  pages = {52--77},
  author = {Davis, Peter Alexander Earls}
}

@book{DeNardis:2020ts,
  title = {The Internet in Everything},
  year = 2020,
  series = {Yale University Press},
  publisher = {Yale University Press},
  urldate = {0},
  author = {DeNardis, Laura}
}

@article{Denning.1994ke,
  title = {The {{US}} Key Escrow Encryption Technology},
  year = 1994,
  journal = {Computer Communications},
  volume = {17},
  number = {7},
  pages = {453--457},
  doi = {10.1016/0140-3664(94)90099-x},
  author = {Denning, Dorothy E.}
}

@article{Diffie:1976dc,
  title = {New Directions in Cryptography},
  year = 1976,
  journal = {IEEE Transactions on Information Theory},
  volume = {22},
  number = {6},
  pages = {644--654},
  doi = {10.1109/tit.1976.1055638},
  author = {Diffie, Whitfield and Hellman, Martin E.}
}

@book{Diffie:2010vm,
  title = {Privacy on the Line},
  year = 2007,
  edition = {2},
  publisher = {MIT},
  address = {Cambridge (MA) \& London},
  author = {Diffie, Whitfield and Landau, Susan}
}

@incollection{Diffie.2007ec,
  title = {The Export of Cryptography in the 20th and the 21st Centuries},
  shorttitle = {The Export of Cryptography},
  booktitle = {The History of Information Security: A Comprehensive Handbook},
  year = 2006,
  pages = {725--769},
  publisher = {Elsevier},
  address = {Amsterdam},
  
  author = {Diffie, Whitfield and Landau, Susan},
  editor = {de Leeuw, Karl and Bergstra, Jan}
}

@article{Duan.2024cm,
  title = {Content Moderation on End-to-End Encrypted Systems: A Legal Analysis},
  shorttitle = {Content Moderation on End-to-End Encrypted Systems},
  year = 2024,
  journal = {Georgetown Law Technology Review},
  volume = {8},
  pages = {1},
  urldate = {2026-01-23},
  author = {Duan, Charles and Grimmelmann, James}
}

@book{Dworkin:1999uv,
  title = {Freedom's Law},
  year = 1999,
  publisher = {Oxford University Press},
  address = {Oxford},
  urldate = {0},
  author = {Dworkin, Ronald}
}

@inbook{Dworkin:1999wn,
  title = {Why Must Speech Be Free?},
  booktitle = {Freedom's Law},
  year = 1999,
  pages = {193--213},
  publisher = {Oxford University Press},
  address = {Oxford},
  author = {Dworkin, Ronald},
  collaborator = {Dworkin, Ronald}
}

@book{Dworkin:2006wo,
  title = {Justice in Robes},
  year = 2006,
  publisher = {Harvard University Press},
  address = {Cambridge (MA) \& London},
  urldate = {0},
  author = {Dworkin, Ronald}
}

@book{Dworkin:2011tm,
  title = {Justice for Hedgehogs},
  year = 2011,
  month = jan,
  publisher = {Harvard University Press},
  address = {Cambridge (MA) \& London},
  urldate = {0},
  author = {Dworkin, Ronald}
}

@misc{EncryptionWorkingGroup:2019wr,
  title = {Moving the Encryption Policy Conversation Forward},
  year = 2019,
  month = oct,
  howpublished = {Carnegie Endowment for International Peace},
  author = {{Encryption Working Group}},
  note = {Retrieved from https://carnegie-production-assets.s3.amazonaws.com/static/files/EWG\_\_Encryption\_Policy.pdf},
  
}

@book{EPP.1997bp,
  title = {The Electronic Privacy Papers: Documents on the Battle for Privacy in the Age of Surveillance},
  shorttitle = {The Electronic Privacy Papers},
  year = 1997,
  publisher = {Wiley},
  address = {New York},
  
  editor = {Schneier, Bruce and Banisar, David}
}

@article{Farid.2022ph,
  title = {An Overview of Perceptual Hashing},
  year = 2021,
  journal = {Journal of Online Trust and Safety},
  volume = {1},
  number = {1},
  doi = {10.54501/jots.v1i1.24},
  
  author = {Farid, Hany}
}

@misc{FBI.2014gd,
  title = {Going Dark: Are Technology, Privacy, and Public Safety on a Collision Course?},
  year = 2014,
  journal = {FBI},
  urldate = {2022-05-10},
  howpublished = {https://archives.fbi.gov/archives/news/speeches/going-dark-are-technology-privacy-and-public-safety-on-a-collision-course},
  author = {Comey, James},
  month = {October 16,},
  note = {Retrieved from https://archives.fbi.gov/archives/news/speeches/going-dark-are-technology-privacy-and-public-safety-on-a-collision-course}
}

@incollection{Feigenbaum:2014el,
  title = {On the Feasibility of a Technological Response to the Surveillance Morass},
  booktitle = {Security {{Protocols XXII}}},
  year = 2014,
  series = {Lecture {{Notes}} in {{Computer Science}}},
  number = {8809},
  pages = {239--252},
  publisher = {Springer International Publishing},
  address = {Cham},
  doi = {10.1007/978-3-319-12400-1_23},
  urldate = {0},
  author = {Feigenbaum, Joan and Koenig, J.{\'e}r{\'e}mie},
  editor = {Christianson, Bruce and Malcolm, James and Maty{\'a}{\v s}, Vashek and {\v S}venda, Petr and Stajano, Frank and Anderson, Jonathan}
}

@incollection{Feigenbaum:2017hg,
  title = {Multiple Objectives of Lawful-Surveillance Protocols ({{Transcript}} of Discussions)},
  booktitle = {Security Protocols {{XXV}}},
  year = 2017,
  series = {Lecture {{Notes}} in {{Computer Science}}},
  number = {10476},
  pages = {9--17},
  publisher = {Springer International Publishing},
  address = {Cham},
  doi = {10.1007/978-3-319-71075-4_1},
  urldate = {0},
  author = {Feigenbaum, Joan and Ford, Bryan},
  editor = {Stajano, Frank and Anderson, Jonathan and Christianson, Bruce and Maty{\'a}{\v s}, Vashek}
}

@incollection{Feigenbaum:2017ur,
  title = {Multiple Objectives of Lawful-Surveillance Protocols},
  booktitle = {Security {{Protocols XXV}}},
  year = 2017,
  series = {Lecture {{Notes}} in {{Computer Science}}},
  number = {10476},
  pages = {1--8},
  publisher = {Springer International Publishing},
  address = {Cham},
  doi = {10.1007/978-3-319-71075-4_1},
  urldate = {0},
  author = {Feigenbaum, Joan and Ford, Bryan},
  editor = {Stajano, Frank and Anderson, Jonathan and Christianson, Bruce and Maty{\'a}{\v s}, Vashek}
}

@article{Feigenbaum:2019fy,
  title = {Encryption and Surveillance: Why the Law-Enforcement Access Question Will Not Just Go Away},
  shorttitle = {Encryption and Surveillance},
  year = 2019,
  month = apr,
  journal = {Communications of the ACM},
  volume = {62},
  number = {5},
  pages = {27--29},
  doi = {10.1145/3319079},
  author = {Feigenbaum, Joan}
}

@article{Fidler.2024zp,
  title = {Zero Progress on Zero-Days: How the Last Ten Years Created the Modern Spyware Market},
  shorttitle = {Zero Progress on Zero-Days},
  year = 2024,
  journal = {Nebraska Law Review},
  volume = {102},
  pages = {713},
  author = {Fidler, Mailyn}
}

@techreport{Finklea.2017lh,
  title = {Law Enforcement Using and Disclosing Technology Vulnerabilities},
  year = 2017,
  number = {R44827},
  address = {Washington, DC},
  institution = {Congressional Research Service},
  
  author = {Finklea, Kirstin},
  type = {Report}
}

@misc{Forsyth.2021ap,
  title = {Apple's {{CSAM}} Detection Technology},
  year = 2021,
  urldate = {2022-05-10},
  author = {Forsyth, David},
  month = {July 13,},
  note = {Retrieved from https://web.archive.org/web/20210805192137/https://www.apple.com/child-safety/pdf/Technical\_Assessment\_of\_CSAM\_Detection\_David\_Forsyth.pdf}
}

@article{Froomkin:1995uz,
  title = {Metaphor Is the Key: Cryptography, the {{Clipper Chip}}, and the {{Constitution}}},
  year = 1995,
  journal = {University of Pennsylvania Law Review},
  volume = {143},
  pages = {709--897},
  author = {Froomkin, A. Michael}
}

@article{Froomkin:1996vm,
  title = {It Came from Planet {{Clipper}}: The Battle over Cryptographic Key ``Escrow''},
  year = 1996,
  journal = {The University of Chicago Legal Forum},
  volume = {1996},
  number = {1},
  pages = {15--75},
  author = {Froomkin, A. Michael}
}

@article{Froomkin:2006ty,
  title = {A Dispatch from the Crypto Wars},
  year = 2006,
  journal = {I/S: A Journal of Law and Policy for the Information Society},
  volume = {2},
  number = {2},
  pages = {345--363},
  author = {Froomkin, A. Michael}
}

@unpublished{Fruhwirth.2005de,
  title = {New Methods in Hard Disk Encryption},
  year = 2005,
  
  author = {Fruhwirth, Clemens},
  month = {July 18,},
  note = {Retrieved from https://clemens.endorphin.org/nmihde/nmihde-A4-os.pdf}
}

@misc{Gill:2018us,
  title = {Shining a Light on the Encryption Debate: A {{Canadian}} Field Guide},
  year = 2018,
  howpublished = {{Citizen Lab \& Samuelson-Glushko Canadian Internet Policy and Public Interest Clinic}},
  author = {Gill, Lex and Israel, Tamir and Parsons, Christopher},
  note = {Retrieved from https://citizenlab.ca/2018/05/shining-light-on-encryption-debate-canadian-field-guide/},
  
}

@misc{Green.2018ro,
  title = {A Few Thoughts on {{Ray Ozzie}}'s ``{{Clear}}'' Proposal},
  year = 2018,
  journal = {A Few Thoughts on Cryptographic Engineering},
  urldate = {2022-05-10},
  author = {Green, Matthew},
  month = {August 26,},
  note = {Retrieved from https://blog.cryptographyengineering.com/2018/04/26/a-few-thoughts-on-ray-ozzies-clear-proposal/},
  howpublished = {A Few Thoughts on Cryptographic Engineering}
}

@inproceedings{Green.2021ar,
  title = {Abuse Resistant Law Enforcement Access Systems},
  booktitle = {Advances in Cryptology -- {{EUROCRYPT}} 2021},
  year = 2021,
  series = {Lecture Notes in Computer Science},
  volume = {12698},
  pages = {553--583},
  publisher = {Springer},
  address = {Cham},
  doi = {10.1007/978-3-030-77883-5_19},
  urldate = {2021},
  
  author = {Green, Matthew and Kaptchuk, Gabriel and Laer, Gijs Van}
}

@article{Gross.2025um,
  title = {{{UK}} Makes New Attempt to Access {{Apple}} Cloud Data},
  year = 2025,
  journal = {Financial Times},
  urldate = {2026-01-26},
  author = {Gross, Anna and Bradshaw, Tim},
  month = {October 1,},
  note = {Retrieved from https://www.ft.com/content/d101fd62-14f9-4f51-beff-ea41e8794265}
}

@book{Gurak.1998,
  title = {Persuasion and Privacy in Cyberspace: The Online Protests over {{Lotus Marketplace}} and the {{Clipper Chip}}},
  year = 1998,
  series = {Contemporary {{Sociology}}},
  publisher = {Yale University Press},
  address = {New Haven, CT \& London},
  doi = {10.2307/2654516},
  
  author = {Kendall, Lori and Gurak, Laura J.}
}

@incollection{Gutwirth:2006up,
  title = {Privacy, Data Protection and Law Enforcement: Opacity of the Individual and Transparency of Power},
  booktitle = {Privacy and the Criminal Law},
  year = 2006,
  pages = {61--104},
  publisher = {Intersentia},
  address = {Antwerp \& Oxford},
  author = {Gutwirth, Serge and Hert, Paul De},
  editor = {Claes, Erik and Duff, Antony and Gutwirth, Serge}
}

@article{Haber.2024lt,
  title = {The Law of the {{Trojan}} Horse},
  year = 2024,
  journal = {U.C. Davis Law Review},
  volume = {57},
  pages = {1667},
  author = {Haber, Eldar}
}

@article{Hardy.2020,
  title = {Australia's Encryption Laws: Practical Need or Political Strategy?},
  year = 2020,
  journal = {Internet Policy Review},
  volume = {9},
  number = {3},
  doi = {10.14763/2020.3.1493},
  
  author = {Hardy, Keiran}
}

@article{Hartzog.2021wp,
  title = {What Is Privacy? {{That}}'s the Wrong Question},
  year = 2021,
  journal = {The University of Chicago Law Review},
  volume = {88},
  number = {1},
  pages = {1677--1688},
  
  author = {Hartzog, Woodrow}
}

@article{Henderson.2013rh,
  title = {Real-Time and Historic Location Surveillance after {{{\emph{United States}}}}{\emph{ v. }}{{{\emph{Jones}}}}: An Administrable, Mildly Mosaic Approach},
  year = 2013,
  journal = {The Journal of Criminal Law \& Criminology},
  volume = {103},
  number = {3},
  pages = {803},
  author = {Henderson, Stephen E.}
}

@article{Hewson.2022td,
  title = {Talking in the Dark: Rules to Facilitate Open Debate about Lawful Access to Strongly Encrypted Information},
  year = 2021,
  journal = {Computer Law \& Security Review},
  volume = {40},
  doi = {10.1016/j.clsr.2020.105526},
  author = {Hewson, Eloise C. and Harrison, Peter S.}
}

@article{Hurwitz.2016ec,
  title = {Encryption{{{\textsuperscript{Congress}}}} Mod ({{Apple}} \& {{CALEA}})},
  year = 2017,
  journal = {Harvard Journal of Law \& Technology},
  volume = {30},
  number = {2},
  pages = {355},
  urldate = {2026-01-23},
  author = {Hurwitz, Justin (Gus)}
}

@article{Jarvis.2021,
  title = {Cypherpunk Ideology: Objectives, Profiles, and Influences (1992--1998)},
  year = 2021,
  journal = {Internet Histories},
  volume = {6},
  number = {3},
  pages = {315--342},
  doi = {10.1080/24701475.2021.1935547},
  
  author = {Jarvis, Craig}
}

@book{Jarvis.2021cw,
  title = {Crypto Wars: The Fight for Privacy in the Digital Age --- a Political History of Digital Encryption},
  shorttitle = {Crypto Wars},
  year = 2021,
  publisher = {CRC Press},
  address = {Boca Raton},
  
  author = {Jarvis, Craig}
}

@book{Kahn:1967tk,
  title = {The Codebreakers},
  year = 1967,
  series = {{{MacMillan}}},
  publisher = {MacMillan},
  urldate = {0},
  author = {Kahn, David}
}

@article{Keenan.2020uk,
  title = {State Access to Encrypted Data in the {{United Kingdom}}: {{The}} `Transparent' Approach},
  year = 2020,
  journal = {Common Law World Review},
  volume = {49},
  number = {3-4},
  pages = {223--244},
  doi = {10.1177/1473779519892641},
  
  author = {Keenan, Bernard}
}

@article{Keenan.2025ii,
  title = {From Interception to Integration: Encryption, Bulk Data, and the Investigatory Powers Regime},
  shorttitle = {From Interception to Integration},
  year = 2025,
  journal = {King's Law Journal},
  volume = {36},
  number = {3},
  pages = {508--539},
  doi = {10.1080/09615768.2025.2551419},
  urldate = {2026-01-26},
  author = {Keenan, Bernard},
  month = {September 2,}
}

@misc{Kehl:2015vz,
  title = {Doomed to Repeat History? {{Lessons}} from the Crypto Wars of the 1990s},
  year = 2015,
  howpublished = {New America},
  author = {Kehl, Danielle and Wilson, Andi and Bankston, Kevin},
  month = {June 17,},
  note = {Retrieved from https://www.newamerica.org/cybersecurity-initiative/policy-papers/doomed-to-repeat-history-lessons-from-the-crypto-wars-of-the-1990s/},
  
}

@inproceedings{Kerschbaumer.2025sh,
  title = {The State of Https Adoption on the Web},
  booktitle = {Proceedings 2025 {{Workshop}} on {{Measurements}}, {{Attacks}}, and {{Defenses}} for the {{Web}}},
  year = 2025,
  publisher = {Internet Society},
  address = {San Diego, CA, USA},
  doi = {10.14722/madweb.2025.23001},
  urldate = {2026-02-01},
  author = {Kerschbaumer, Christoph and Braun, Frederik and Friedberger, Simon and J{\"u}rgens, Malte}
}

@book{Koops:1999uk,
  title = {The Crypto Controversy},
  year = 1999,
  series = {Kluwer},
  publisher = {Kluwer},
  author = {Koops, Bert-Jaap}
}

@article{Koops:2018bh,
  title = {Looking for Some Light through the Lens of "Cryptowar" History: Policy Options for Law Enforcement Authorities against "Going Dark"},
  year = 2018,
  journal = {Computer Law \& Security Review},
  volume = {34},
  number = {4},
  pages = {890--900},
  doi = {10.1016/j.clsr.2018.06.003},
  author = {Koops, Bert-Jaap and Kosta, Eleni}
}

@book{Landau:2011wy,
  title = {Surveillance or Security?},
  year = 2011,
  series = {{{MIT}} Press},
  publisher = {MIT Press},
  urldate = {0},
  author = {Landau, Susan}
}

@book{Landau:2017uy,
  title = {Listening in: Cybersecurity in an Insecure Age},
  year = 2017,
  month = nov,
  publisher = {Yale University Press},
  address = {New Haven \& London},
  author = {Landau, Susan}
}

@article{Landau.2013ms,
  title = {Making Sense from {{Snowden}}: What's Significant in the {{NSA}} Surveillance Revelations},
  shorttitle = {Making {{Sense}} from {{Snowden}}},
  year = 2013,
  month = jul,
  journal = {IEEE Security \& Privacy},
  volume = {11},
  number = {4},
  pages = {54--63},
  doi = {10.1109/MSP.2013.90},
  urldate = {2026-01-27},
  copyright = {https://ieeexplore.ieee.org/Xplorehelp/downloads/license-information/IEEE.html},
  author = {Landau, Susan}
}

@article{Landau.2014hm,
  title = {Highlights from Making Sense of {{Snowden}}, Part {{II}}: What's Significant in the {{NSA}} Revelations},
  shorttitle = {Highlights from Making Sense of Snowden, Part {{II}}},
  year = 2014,
  month = jan,
  journal = {IEEE Security \& Privacy},
  volume = {12},
  number = {1},
  pages = {62--64},
  doi = {10.1109/MSP.2013.161},
  urldate = {2026-01-27},
  copyright = {https://ieeexplore.ieee.org/Xplorehelp/downloads/license-information/IEEE.html},
  author = {Landau, Susan}
}

@misc{Landau.2016et,
  title = {The Encryption Tightrope: Balancing {{Americans}}' Security and Privacy},
  shorttitle = {Encryption Tightrope},
  year = 2016,
  number = {114-98},
  pages = {104--130},
  author = {Landau, Susan},
  month = {March 1,},
  howpublished = {Hearing before the Committee on the Judiciary, 114th Cong.}
}

@misc{Landau.2018bs,
  title = {Building on Sand Isn't Stable: Correcting a Misunderstanding of the {{National Academies}} Report on Encryption},
  year = 2018,
  journal = {Lawfare},
  author = {Landau, Susan},
  month = {April 25,},
  note = {Retrieved from https://www.lawfareblog.com/building-sand-isnt-stable-correcting-misunderstanding-national-academies-report-encryption},
  howpublished = {Lawfare}
}

@misc{Landau.2018ro,
  title = {What's Involved in Vetting a Security Protocol: Why {{Ray Ozzie}}'s Proposal for Exceptional Access Does Not Pass Muster},
  year = 2018,
  journal = {Lawfare},
  urldate = {2022-05-10},
  author = {Landau, Susan},
  month = {May 14,},
  note = {Retrieved from https://www.lawfareblog.com/whats-involved-vetting-security-protocol-why-ray-ozzies-proposal-exceptional-access-does-not-pass},
  howpublished = {Lawfare}
}

@incollection{Landau.2022dc,
  title = {The Development of a Crypto Policy Community: {{Diffie}}--{{Hellman}}'s Impact on Public Policy},
  shorttitle = {The Development of a Crypto Policy Community},
  booktitle = {Democratizing Cryptography},
  year = 2022,
  edition = {1},
  pages = {213--256},
  publisher = {ACM},
  address = {New York, NY, USA},
  doi = {10.1145/3549993.3550002},
  urldate = {2026-01-16},
  author = {Landau, Susan},
  editor = {Slayton, Rebecca},
  month = {August 24,}
}

@misc{Landau.2022ns,
  title = {Normalizing Surveillance},
  year = 2021,
  journal = {Lawfare},
  urldate = {2022-05-10},
  author = {Landau, Susan},
  month = {August 30,},
  note = {Retrieved from https://www.lawfareblog.com/normalizing-surveillance},
  howpublished = {Lawfare}
}

@article{Levison.2014lb,
  title = {Secrets, Lies and {{Snowden}}'s Email: Why {{I}} Was Forced to Shut down {{Lavabit}}},
  year = 2014,
  month = may,
  journal = {Guardian},
  author = {Levison, Ladar}
}

@book{Levy:2002wp,
  title = {Crypto},
  year = 2002,
  publisher = {Viking},
  urldate = {0},
  author = {Levy, Steven}
}

@article{Levy.2016wa,
  title = {Why Are We Fighting the Crypto Wars Again?},
  year = 2016,
  journal = {Wired},
  urldate = {2026-01-29},
  author = {Levy, Steven},
  month = {March 11,},
  note = {Retrieved from https://www.wired.com/2016/03/why-are-we-fighting-the-crypto-wars-again/}
}

@article{Levy.2018cr,
  title = {Can This System of Unlocking Phones Crack the Crypto War?},
  year = 2018,
  journal = {Wired},
  author = {Levy, Steven},
  month = {April 25,},
  note = {Retrieved from https://www.wired.com/story/crypto-war-clear-encryption/}
}

@article{Li:2018kp,
  title = {A Comprehensive Overview of Government Hacking Worldwide},
  year = 2018,
  month = jan,
  journal = {IEEE access : practical innovations, open solutions},
  volume = {6},
  pages = {55053--55073},
  doi = {10.1109/access.2018.2871762},
  author = {Li, Chen-Yu and Huang, Chien-Cheng and Lai, Feipei and Lee, San-Liang and Wu, Jingshown}
}

@article{LiguoriFilho:2020vq,
  title = {Exploring Lawful Hacking as a Possible Answer to the ``Going Dark'' Debate},
  year = 2020,
  journal = {Michigan Technology Law Review},
  volume = {26},
  number = {2},
  pages = {317--345},
  author = {Liguori Filho, Carlos Augusto}
}

@article{Magal.2003cp,
  title = {The State and Telecom Surveillance Policy: The {{Clipper}} Chip Initiative},
  shorttitle = {The {{Clipper}} Chip Initiative},
  year = 2003,
  journal = {Communication Law and Policy},
  volume = {8},
  number = {4},
  pages = {429--464},
  doi = {10.1207/s15326926clp0804_03},
  
  author = {{Pednekar-Magal}, Vandana and Shields, Peter}
}

@article{Markoff.1994fd,
  title = {Flaw Discovered in Federal Plan for Wiretapping},
  year = 1994,
  journal = {New York Times},
  pages = {A1},
  urldate = {2025-09-30},
  author = {Markoff, John},
  month = {June 2,},
  note = {Retrieved from https://www.nytimes.com/1994/06/02/us/flaw-discovered-in-federal-plan-for-wiretapping.html}
}

@article{Mayer:2018vc,
  title = {Government Hacking},
  year = 2018,
  journal = {The Yale Law Journal},
  volume = {127},
  number = {3},
  pages = {570--662},
  author = {Mayer, Jonathan}
}

@article{McConnell:2015vu,
  title = {Why the Fear over Ubiquitous Data Encryption Is Overblown},
  year = 2015,
  journal = {Washington Post},
  author = {McConnell, Mike and Chertoff, Michael and Lynn, William},
  month = {July 28,},
  note = {Retrieved from https://www.washingtonpost.com/opinions/the-need-for-ubiquitous-data-encryption/2015/07/28/3d145952-324e-11e5-8353-1215475949f4\_story.html}
}

@article{Menn.2025uo,
  title = {U.{{K}}. Orders {{Apple}} to Let It Spy on Users' Encrypted Accounts},
  year = 2025,
  journal = {The Washington Post},
  urldate = {2026-01-29},
  author = {Menn, Joseph},
  month = {February 7,},
  note = {Retrieved from https://www.washingtonpost.com/technology/2025/02/07/apple-encryption-backdoor-uk/}
}

@misc{Miller.2025uh,
  title = {{{UK}} Has `Agreed to Drop' Demand for Access to {{Apple}} User Data, Says {{US}}},
  year = 2025,
  urldate = {2026-01-29},
  howpublished = {https://www.ft.com/content/ab0aba27-81e0-4ee5-bcbb-6bce85386e40},
  author = {Miller, Joe and Bradshaw, Tim and Gross, Anna and Parker, George},
  month = {August 19,},
  note = {Retrieved from https://www.ft.com/content/ab0aba27-81e0-4ee5-bcbb-6bce85386e40}
}

@misc{Mohanty.2019ei,
  title = {The Encryption Debate in {{India}}},
  year = 2019,
  address = {Washington, DC},
  howpublished = {Carnegie Endowment for International Peace},
  
  author = {Mohanty, Bedavyasa},
  month = {May 30,},
  note = {Retrieved from https://carnegieendowment.org/posts/2019/05/the-encryption-debate-in-india},
  
}

@misc{Mohanty.2021ei,
  title = {The Encryption Debate in {{India}}: 2021 Update},
  year = 2021,
  address = {Washington, DC},
  howpublished = {Carnegie Endowment for International Peace},
  
  author = {Mohanty, Bedavyasa},
  month = {March 31,},
  note = {Retrieved from https://carnegieendowment.org/posts/2021/03/the-encryption-debate-in-india-2021-update},
  
}

@book{Moore:2010tm,
  title = {Privacy Rights: Moral and Legal Foundations},
  year = 2010,
  publisher = {Pennsylvania State University},
  author = {Moore, Adam D.}
}

@article{Narayanan:2013ij,
  title = {What Happened to the Crypto Dream? {{Part}} 2},
  year = 2013,
  journal = {IEEE Security \& Privacy},
  volume = {11},
  number = {3},
  pages = {68--71},
  doi = {10.1109/msp.2013.75},
  author = {Narayanan, Arvind}
}

@book{NationalAcademiesofSciencesEngineeringandMedicine:2018br,
  title = {Decrypting the Encryption Debate: A Framework for Decision Makers},
  year = 2018,
  series = {National Academies Press},
  publisher = {National Academies Press},
  doi = {10.17226/25010},
  urldate = {0},
  author = {{National Academies of Sciences, Engineering and Medicine}}
}

@phdthesis{Nguyen:2017wp,
  type = {Master's Thesis ({{Security Studies}})},
  title = {Lawful Hacking: Toward a Middle-Ground Solution to the Going Dark Problem},
  year = 2017,
  school = {Naval Postgraduate School},
  author = {Nguyen, Hoaithi Y. T.}
}

@article{Nissenbaum.2005wc,
  title = {Where Computer Security Meets National Security},
  year = 2005,
  month = jun,
  journal = {Ethics and Information Technology},
  volume = {7},
  number = {2},
  pages = {61--73},
  doi = {10.1007/s10676-005-4582-3},
  urldate = {2026-01-16},
  copyright = {http://www.springer.com/tdm},
  author = {Nissenbaum, Helen}
}

@book{NRC.1991cr,
  title = {Computers at Risk: Safe Computing in the Information Age},
  year = 1991,
  publisher = {National Academy Press},
  address = {Washington, DC},
  doi = {10.17226/1581},
  
  author = {{National Research Council}}
}

@book{NRC.1996,
  title = {Cryptography's Role in Securing the Information Society},
  year = 1996,
  publisher = {The National Academies Press},
  address = {Washington, DC},
  doi = {10.17226/5131},
  
  editor = {Dam, Kenneth W. and Lin, Herbert S.}
}

@article{NYT.2014ps,
  title = {Signaling Post-{{Snowden}} Era, New {{iPhone}} Locks out {{N}}.{{S}}.{{A}}},
  year = 2014,
  journal = {New York Times},
  pages = {A1},
  author = {Sanger, David E. and Chen, Brian X.},
  month = {September 26,},
  note = {Retrieved from https://www.nytimes.com/2014/09/27/technology/iphone-locks-out-the-nsa-signaling-a-post-snowden-era-.html}
}

@article{NYT.2016ap,
  title = {U.{{S}}. Says It Has Unlocked {{iPhone}} without {{Apple}}},
  year = 2016,
  journal = {New York Times},
  pages = {A1},
  author = {Benner, Katie and Lichtblau, Eric},
  month = {March 28,},
  note = {Retrieved from https://www.nytimes.com/2016/03/29/technology/apple-iphone-fbi-justice-department-case.html}
}

@article{NYT.2019ca,
  title = {Child Abusers Run Rampant as Tech Companies Look the Other Way},
  year = 2019,
  journal = {New York Times},
  author = {Keller, Michael H. and Dance, Gabriel J. X.},
  month = {November 9,},
  note = {Retrieved from https://www.nytimes.com/interactive/2019/11/09/us/internet-child-sex-abuse.html}
}

@article{NYT.2020ae,
  title = {As {{Apple}} Resists, Encryption Fray Erupts in Battle},
  year = 2016,
  journal = {The New York Times},
  pages = {A1},
  address = {New York, NY},
  author = {Lichtblau, Eric and Benner, Katie},
  month = {February 18,},
  note = {Retrieved from https://www.nytimes.com/2016/02/18/technology/apple-timothy-cook-fbi-san-bernardino.html}
}

@article{NYT.2020cs,
  title = {Tech Companies Detect a Surge in Online Videos of Child Sexual Abuse},
  year = 2020,
  journal = {New York Times},
  author = {Dance, Gabriel J. .X. and Keller, Michael H.},
  month = {February 7,},
  note = {Retrieved from https://www.nytimes.com/2020/02/07/us/online-child-sexual-abuse.html}
}

@article{ONeill.2017ti,
  title = {{{TLS}} Inspection: How Often and Who Cares?},
  shorttitle = {{{TLS}} Inspection},
  year = 2017,
  journal = {IEEE Internet Computing},
  volume = {21},
  number = {3},
  pages = {22--29},
  doi = {10.1109/MIC.2017.58},
  urldate = {2026-02-01},
  author = {O'Neill, Mark and Ruoti, Scott and Seamons, Kent and Zappala, Daniel},
  month = {May 1,}
}

@misc{OTI.2017uk,
  title = {Deciphering the {{European}} Encryption Debate: {{United Kingdom}}},
  year = 2017,
  month = jun,
  address = {Washington, DC},
  howpublished = {New America},
  
  author = {Acharya, Bhairav and Bankston, Kevin and Schulman, Ross and Wilson, Andi},
  note = {Retrieved from https://www.newamerica.org/oti/policy-papers/deciphering-european-encryption-debate-united-kingdom/},
  
}

@article{Owsley.2017lb,
  title = {Lavabitten},
  year = 2017,
  journal = {West Virginia Law Review},
  volume = {119},
  number = {3},
  pages = {941--956},
  
  author = {Owsley, Brian L.}
}

@misc{Ozzie.2018cr,
  title = {{{CLEAR}}},
  year = 2018,
  urldate = {2022-05-10},
  
  author = {Ozzie, Ray},
  note = {Retrieved from https://github.com/rayozzie/clear/blob/master/clear-rozzie.pdf}
}

@article{Pell:2016wm,
  title = {You Can't Always Get What You Want: How Will Law Enforcement Get What It Needs in a Post-{{CALEA}}, Cybersecurity-Centric Encryption Era},
  year = 2016,
  journal = {North Carolina Journal of Law Technology},
  volume = {17},
  number = {4},
  pages = {599--644},
  author = {Pell, Stephanie K.}
}

@article{Perlroth.2013na,
  title = {N.{{S}}.{{A}}. Able to Foil Basic Safeguards of Privacy on Web},
  year = 2013,
  journal = {The New York Times},
  urldate = {2026-01-29},
  chapter = {U.S.},
  author = {Perlroth, Nicole and Larson, Jeff and Shane, Scott},
  month = {September 5,},
  note = {Retrieved from https://www.nytimes.com/2013/09/06/us/nsa-foils-much-internet-encryption.html}
}

@article{Pfefferkorn:2017vd,
  title = {Everything Radiates: Does the {{Fourth Amendment}} Regulate Side-Channel Cryptanalysis?},
  year = 2017,
  journal = {Connecticut Law Review},
  volume = {49},
  number = {5},
  pages = {1393--1452},
  author = {Pfefferkorn, Riana}
}

@misc{Pfefferkorn:2018uf,
  title = {The Risks of "Responsible Encryption"},
  year = 2018,
  month = feb,
  howpublished = {{Center for Internet and Society, Stanford Law School}},
  urldate = {0},
  author = {Pfefferkorn, Riana},
  note = {Retrieved from https://cyberlaw.stanford.edu/publications/risks-responsible-encryption},
  
}

@misc{Pfefferkorn:2018wx,
  title = {Security Risks of Government Hacking},
  year = 2018,
  howpublished = {{Center for Internet and Society, Stanford Law School}},
  
  author = {Pfefferkorn, Riana},
  month = {September 5,},
  note = {Retrieved from https://cyberlaw.stanford.edu/publications/security-risks-government-hacking/},
  
}

@misc{Pinkas.2021ap,
  title = {A Review of the Cryptography behind the {{Apple PSI}} System},
  year = 2021,
  author = {Pinkas, Benny},
  month = {July 9,},
  note = {Retrieved from https://web.archive.org/web/20210805190856/https://www.apple.com/child-safety/pdf/Technical\_Assessment\_of\_CSAM\_Detection\_Benny\_Pinkas.pdf}
}

@misc{Podchasov.2024hr,
  year = 2024,
  month = {February 3,},
  note = {Retrieved from https://hudoc.echr.coe.int/?i=001-230854},
  title = {Podchasov v. {{Russia}} ({{Application}} No. 33696/19)},
  key = {European Court of Human Rights (Third Section)},
  howpublished = {European Court of Human Rights (Third Section)}
}

@article{Pozen.2016pp,
  title = {Privacy--Privacy Tradeoffs},
  year = 2016,
  journal = {University of Chicago Law Review},
  volume = {83},
  number = {1},
  pages = {221--247},
  
  author = {Pozen, David E.}
}

@article{Queiroz.2022,
  title = {Cypherpunk},
  year = 2022,
  journal = {Internet Policy Review},
  volume = {11},
  number = {2},
  doi = {10.14763/2022.2.1664},
  
  author = {Ramiro, Andr{\'e} and de Queiroz, Ruy}
}

@book{Quinlan:2016wr,
  title = {A Brief History of Law Enforcement Hacking in the {{United States}}},
  year = 2016,
  publisher = {New America},
  urldate = {0},
  author = {Quinlan, Sayako and Wilson, Andi}
}

@article{Reuters.2020af,
  title = {Apple Dropped Plan for Encrypting Backups after {{FBI}} Complained},
  year = 2020,
  journal = {Reuters},
  author = {Menn, Joseph},
  month = {January 21,},
  note = {Retrieved from https://www.reuters.com/article/us-apple-fbi-icloud-exclusive/exclusive-apple-dropped-plan-for-encrypting-backups-after-fbi-complained-sources-idUSKBN1ZK1CT}
}

@book{Richards.2022wp,
  title = {Why Privacy Matters},
  year = 2022,
  publisher = {Oxford University Press},
  address = {Oxford},
  
  author = {Richards, Neil M.}
}

@misc{Rogaway:2015uc,
  title = {The Moral Character of Cryptographic Work},
  year = 2015,
  number = {2015/1162},
  eprint = {2015/1162},
  publisher = {Cryptology Print Archive},
  archiveprefix = {Cryptology Print Archive},
  author = {Rogaway, Phillip},
  note = {Retrieved from http://eprint.iacr.org/2015/1162}
}

@misc{Rogers.2016ce,
  title = {Encryption and Cyber Matters},
  year = 2016,
  number = {114--671},
  pages = {47--52},
  author = {Lettre, Marcell J. and Rogers, Michael S.},
  month = {September 13,},
  howpublished = {Hearing before the Committee on Armed Services United States Senate, 114th Cong.}
}

@misc{Rosenstein.2017ec,
  title = {Deputy {{Attorney General Rod J}}. {{Rosenstein}} Delivers Remarks on Encryption at the {{United States Naval Academy}}},
  year = 2017,
  urldate = {2022-05-10},
  author = {{U.S. Department of Justice}},
  month = {October 10,},
  note = {Retrieved from https://www.justice.gov/opa/speech/deputy-attorney-general-rod-j-rosenstein-delivers-remarks-encryption-united-states-naval}
}

@article{Rozenshtein:2018wq,
  title = {Surveillance Intermediaries},
  year = 2018,
  journal = {Stanford Law Review},
  volume = {70},
  number = {1},
  pages = {99--189},
  author = {Rozenshtein, Alan Z.}
}

@article{Rozenshtein:2019vf,
  title = {Wicked Crypto},
  year = 2019,
  journal = {UC Irvine Law Review},
  volume = {9},
  pages = {1181--1216},
  author = {Rozenshtein, Alan Z.}
}

@inproceedings{Savage:2018cd,
  title = {Lawful Device Access without Mass Surveillance Risk},
  booktitle = {Proceedings of the 2018 {{ACM SIGSAC Conference}} on {{Computer}} and {{Communications Security}}},
  year = 2018,
  pages = {1761--1774},
  doi = {10.1145/3243734.3243758},
  author = {Savage, Stefan}
}

@inproceedings{Scheffler.2023pv,
  title = {Public Verification for Private Hash Matching},
  booktitle = {2023 {{IEEE Symposium}} on {{Security}} and {{Privacy}} ({{SP}})},
  year = 2023,
  month = may,
  pages = {253--273},
  publisher = {IEEE},
  address = {San Francisco, CA, USA},
  doi = {10.1109/SP46215.2023.10179349},
  urldate = {2025-11-07},
  copyright = {https://doi.org/10.15223/policy-009},
  author = {Scheffler, Sarah and Kulshrestha, Anunay and Mayer, Jonathan}
}

@article{Scheffler.2023sc,
  title = {{{SoK}}: {{Content}} Moderation for End-to-End Encryption},
  year = 2023,
  journal = {Proceedings on Privacy Enhancing Technologies},
  volume = {2023},
  number = {2},
  eprint = {2303.03979},
  pages = {403--429},
  doi = {10.56553/popets-2023-0060},
  
  author = {Scheffler, Sarah and Mayer, Jonathan}
}

@incollection{Schneier.1997fb,
  title = {The Field of Battle: An Overview},
  booktitle = {The Electronic Privacy Papers: Documents on the Battle for Privacy in the Age of Surveillance},
  year = 1997,
  pages = {291--338},
  publisher = {Wiley},
  address = {New York},
  
  author = {Schneier, Bruce and Banisar, David},
  editor = {Schneier, Bruce and Banisar, David}
}

@inproceedings{Segal:2014tl,
  title = {Catching Bandits and Only Bandits: Privacy-Preserving Intersection Warrants for Lawful Surveillance},
  booktitle = {{{IV FOCI}}},
  year = 2014,
  month = jul,
  author = {Segal, Aaron and Ford, Bryan and Feigenbaum, Joan}
}

@misc{Segal:2016bh,
  title = {Open, Privacy-Preserving Protocols for Lawful Surveillance},
  year = 2016,
  number = {1607.03659},
  eprint = {1607.03659},
  publisher = {arXiv},
  doi = {10.48550/arXiv.1607.03659},
  archiveprefix = {arXiv},
  author = {Segal, Aaron and Feigenbaum, Joan and Ford, Bryan},
  month = {July 13,}
}

@misc{ServanSchreiber:2020ts,
  title = {Judge, Jury \& Encryptioner: Exceptional Device Access with a Social Cost},
  year = 2020,
  number = {1912.05620},
  eprint = {1912.05620},
  publisher = {arXiv},
  doi = {10.48550/arXiv.1912.05620},
  archiveprefix = {arXiv},
  author = {{Servan-Schreiber}, Sacha and Wheeler, Archer},
  month = {March 6,}
}

@article{Shapley.1977cs,
  title = {Cryptology: Scientists Puzzle over Threat to Open Research, Publication},
  shorttitle = {Cryptology},
  year = 1977,
  journal = {Science},
  volume = {197},
  number = {4311},
  pages = {1345--1349},
  doi = {10.1126/science.197.4311.1345},
  urldate = {2026-01-26},
  author = {Shapley, Deborah and Kolata, Gina Bari},
  month = {September 30,}
}

@article{Sherman.2023cw,
  title = {{\emph{The }}{{{\emph{Codebreakers}}}} War: {{David Kahn}}, {{Macmillan}}, the Government, and the Making of a Cryptologic History Masterpiece},
  shorttitle = {{\emph{The }}{{{\emph{Codebreakers}}}} War},
  year = 2023,
  journal = {Cryptologia},
  volume = {47},
  number = {3},
  pages = {205--226},
  doi = {10.1080/01611194.2021.1998808},
  urldate = {2026-01-26},
  author = {Sherman, David},
  month = {May 4,}
}

@article{Shurson.2024er,
  title = {A {{European}} Right to End-to-End Encryption?},
  year = 2024,
  month = nov,
  journal = {Computer Law \& Security Review},
  volume = {55},
  pages = {106063},
  doi = {10.1016/j.clsr.2024.106063},
  urldate = {2025-05-17},
  author = {Shurson, Jessica}
}

@misc{Silva.2021eb,
  title = {The Encryption Debate in {{Brazil}}: 2021 Update},
  year = 2021,
  address = {Washington, DC},
  howpublished = {Carnegie Endowment for International Peace},
  
  author = {Silva, Priscilla and Mangeth, Ana Lara and Perrone, Christian},
  month = {March 31,},
  note = {Retrieved from https://carnegieendowment.org/posts/2021/03/the-encryption-debate-in-brazil-2021-update},
  
}

@incollection{SKIPJACK.1995cl,
  title = {{{SKIPJACK}} Review: Interim Report},
  booktitle = {Building in {{Big Brother}}: The Cryptographic Policy Debate},
  year = 1995,
  pages = {119--130},
  publisher = {Springer},
  address = {New York},
  author = {Brickell, Ernest F. and Denning, Dorothy E. and Kent, Stephen T. and Maher, David P. and Tuchman, Walter},
  editor = {Hoffman, Lance J.}
}

@article{Solove:2002gr,
  title = {Conceptualizing Privacy},
  year = 2002,
  journal = {California Law Review},
  volume = {90},
  pages = {1087--1155},
  author = {Solove, Daniel J.}
}

@article{Solove:2007up,
  title = {``{{I}}'ve Got Nothing to Hide,'' and Other Misunderstandings of Privacy},
  year = 2007,
  journal = {San Diego Law Review},
  volume = {44},
  number = {4},
  pages = {745--772},
  author = {Solove, Daniel J.}
}

@article{Solove:2010vr,
  title = {Fourth {{Amendment}} Pragmatism},
  year = 2010,
  journal = {Boston College Law Review},
  volume = {51},
  pages = {1511--1538},
  author = {Solove, Daniel J.}
}

@book{Solove:2011vz,
  title = {Nothing to Hide: {{The}} False Tradeoff between Privacy and Security},
  year = 2011,
  publisher = {Yale University Press},
  urldate = {0},
  author = {Solove, Daniel J.}
}

@book{Solove.2008up,
  title = {Understanding Privacy},
  year = 2008,
  publisher = {Harvard University Press},
  address = {Cambridge, MA \& London},
  author = {Solove, Daniel J.}
}

@article{Sparkes.2021pf,
  title = {Possible Flaw in Protection Algorithm},
  year = 2021,
  journal = {New Scientist},
  volume = {251},
  number = {3349},
  pages = {8},
  doi = {10.1016/s0262-4079(21)01484-6},
  author = {Sparkes, Matthew},
  month = {August 28,}
}

@article{Stamos.2021ap,
  title = {Apple Wants to Protect Children. but It's Creating Serious Privacy Risks},
  year = 2021,
  journal = {New York Times},
  author = {Green, Matthew and Stamos, Alex},
  month = {August 11,},
  note = {Retrieved from https://www.nytimes.com/2021/08/11/opinion/apple-iphones-privacy.html}
}

@misc{Stilgherrian.2019ea,
  title = {The Encryption Debate in {{Australia}}},
  year = 2019,
  address = {Washington, DC},
  howpublished = {Carnegie Endowment for International Peace},
  
  author = {{Stilgherrian}},
  month = {May 30,},
  note = {Retrieved from https://carnegieendowment.org/posts/2019/05/the-encryption-debate-in-australia},
  
}

@misc{Stilgherrian.2021ea,
  title = {The Encryption Debate in {{Australia}}: 2021 Update},
  year = 2021,
  address = {Washington, DC},
  howpublished = {Carnegie Endowment for International Peace},
  
  author = {{Stilgherrian}},
  month = {March 31,},
  note = {Retrieved from https://carnegieendowment.org/posts/2021/03/the-encryption-debate-in-australia-2021-update},
  
}

@article{Strupekk.2022ph,
  title = {Learning to Break Deep Perceptual Hashing: {{The}} Use Case {{NeuralHash}}},
  year = 2022,
  journal = {2022 ACM Conference on Fairness, Accountability, and Transparency},
  eprint = {2111.06628},
  pages = {58--69},
  doi = {10.1145/3531146.3533073},
  
  author = {Struppek, Lukas and Hintersdorf, Dominik and Neider, Daniel and Kersting, Kristian}
}

@article{Swire.2011eg,
  title = {Encryption and Globalization},
  year = 2011,
  journal = {Columbia Science \& Technology Law Review},
  volume = {23},
  pages = {416--481},
  
  author = {Swire, Peter and Ahmad, Kenesa}
}

@misc{Tromer.2018cr,
  title = {Eran Tromer's Attack on Ray Ozzie's {{CLEAR}} Protocol},
  year = 2018,
  journal = {SMBlog},
  urldate = {2022-05-10},
  author = {Tromer, Eran},
  month = {May 2,},
  note = {Retrieved from https://www.cs.columbia.edu/\textasciitilde smb/blog/2018-05/2018-05-02.html},
  howpublished = {SMBlog}
}

@misc{Tuchtfeld.2022ty,
  title = {"{{Thank}} You Very Much, Your Mail Is Perfectly Fine": How the {{European Commission}} Wants to Abolish the Secrecy of Correspondence in the Digital Sphere},
  shorttitle = {"{{Your}} Mail Is Perfectly Fine"},
  year = 2022,
  journal = {Verfassungsblog},
  doi = {10.17176/20220818-181927-0},
  urldate = {2026-01-16},
  copyright = {Creative Commons Attribution Share Alike 4.0 International},
  author = {Tuchtfeld, Erik},
  month = {August 18,},
  note = {Retrieved from https://intrechtdok.de/receive/mir\_mods\_00013535},
  howpublished = {Verfassungsblog}
}

@misc{UnitedNationsHumanRightsCouncil:2015ta,
  title = {Report of the {{Special Rapporteur}} on the Promotion and Protection of the Right to Freedom of Opinion and Expression, {{David Kaye}} ({{A}}/{{HRC}}/29/32)},
  year = 2015,
  key = {United Nations Human Rights Council},
  howpublished = {United Nations Human Rights Council},
  month = {May 22,}
}

@article{Upson.2021,
  title = {Laws of Encryption: {{An}} Emerging Legal Framework},
  year = 2021,
  journal = {Computer Law \& Security Review},
  volume = {43},
  pages = {1--21},
  doi = {10.1016/j.clsr.2021.105635},
  author = {Dizon, Michael Anthony C. and Upson, Peter John}
}

@misc{Upturn.2020me,
  title = {Mass Extraction: The Widespread Power of u.s. Law Enforcement to Search Mobile Phones},
  shorttitle = {Mass Extraction},
  year = 2020,
  howpublished = {Upturn},
  
  author = {Koepke, Logan and Weil, Emma and Janardan, Urmila and Dada, Tinuola and Yu, Harlan},
  
}

@article{Uresk.2020cd,
  title = {Compelling Suspects to Unlock Their Phones: Recommendations for Prosecutors and Law Enforcement},
  year = 2020,
  journal = {Brigham Young University Law Review},
  volume = {46},
  number = {2},
  pages = {601--656},
  
  author = {Uresk, Carissa A.}
}

@article{USSen.1978de,
  title = {Unclassified Summary: {{Involvement}} of {{NSA}} in the Development of the Data Encryption Standard},
  year = 1978,
  journal = {IEEE Communications Society Magazine},
  volume = {16},
  number = {6},
  pages = {53--55},
  doi = {10.1109/mcom.1978.1089789},
  
  author = {{U.S. Senate Select Committee on Intelligence}}
}

@misc{Varia.2018le,
  title = {A Roadmap for Exceptional Access Research},
  year = 2018,
  journal = {Lawfare},
  author = {Varia, Mayank},
  month = {December 5,},
  note = {Retrieved from https://www.lawfareblog.com/roadmap-exceptional-access-research},
  howpublished = {Lawfare}
}

@article{Vice.2018gs,
  title = {Someone Is Trying to Extort {{iPhone}} Crackers {{Grayshift}} with Leaked Code},
  year = 2018,
  month = apr,
  journal = {Vice},
  author = {Cox, Joseph}
}

@incollection{Vidan.2025cg,
  title = {Cryptography Goes Public: Contesting the Meaning of a New Field in the 1970s {{United States}}},
  booktitle = {Just Code: Power, Inequality, and the Political Economy of {{IT}}},
  year = 2025,
  pages = {372--387},
  publisher = {Johns Hopkins University Press},
  author = {Vidan, Gili},
  editor = {Yost, Jeffrey R. and D{\'i}az, Gerardo}
}

@article{WaPo.2014ae,
  title = {Newest {{Androids}} Will Join {{iPhones}} in Offering Default Encryption, Blocking Police},
  year = 2014,
  journal = {Washington Post},
  chapter = {The Switch},
  author = {Timberg, Craig},
  month = {September 14,},
  note = {Retrieved from https://www.washingtonpost.com/news/the-switch/wp/2014/09/18/newest-androids-will-join-iphones-in-offering-default-encryption-blocking-police/}
}

@article{WaPo.2018oe,
  title = {{{FBI}} Repeatedly Overstated Encryption Threat Figures to {{Congress}}, Public},
  year = 2018,
  journal = {Washington Post},
  author = {Barrett, Devlin},
  month = {May 22,},
  note = {Retrieved from https://www.washingtonpost.com/world/national-security/fbi-repeatedly-overstated-encryption-threat-figures-to-congress-public/2018/05/22/5b68ae90-5dce-11e8-a4a4-c070ef53f315\_story.html}
}

@article{Weigel.2022cs,
  title = {Apple's '{{Communication Safety}}' Feature for Child Users: Implications for Law Enforcement's Ability to Compel {{iMessage}} Decryption},
  year = 2022,
  journal = {Stanford Technology Law Review},
  volume = {24},
  number = {2},
  pages = {210--246},
  
  author = {Weigel, Nicholas A.}
}

@article{Weippi.2016,
  title = {A Systematic Analysis of the {{Juniper Dual EC}} Incident},
  year = 2016,
  journal = {Proceedings of the 2016 ACM SIGSAC Conference on Computer and Communications Security},
  pages = {468--479},
  doi = {10.1145/2976749.2978395},
  
  author = {Weippl, Edgar and Katzenbeisser, Stefan and Kruegel, Christopher and Myers, Andrew and Halevi, Shai and Checkoway, Stephen and Maskiewicz, Jacob and Garman, Christina and Fried, Joshua and Cohney, Shaanan and Green, Matthew and Heninger, Nadia and Weinmann, Ralf-Philipp and Rescorla, Eric and Shacham, Hovav}
}

@article{West.2022ci,
  title = {Cryptography as Information Control},
  year = 2022,
  month = jun,
  journal = {Social Studies of Science},
  volume = {52},
  number = {3},
  pages = {353--375},
  doi = {10.1177/03063127221078314},
  urldate = {2026-01-25},
  author = {West, Sarah Myers}
}

@article{Wired.2014wa,
  title = {{{WhatsApp}} Just Switched on End-to-End Encryption for Hundreds of Millions of Users},
  year = 2014,
  journal = {Wired},
  author = {Greenberg, Andy},
  month = {November 18,},
  note = {Retrieved from https://www.wired.com/2014/11/whatsapp-encrypted-messaging/}
}

@article{Wired.2021at,
  title = {Apple Walks a Privacy Tightrope to Spot Child Abuse in {{iCloud}}},
  year = 2021,
  journal = {Wired},
  author = {Greenberg, Andy},
  month = {August 5,},
  note = {Retrieved from https://www.wired.com/story/apple-csam-detection-icloud-photos-encryption-privacy/}
}

@inproceedings{Wright:2018dj,
  title = {Crypto Crumple Zones: Enabling Limited Access without Mass Surveillance},
  shorttitle = {Crypto Crumple Zones},
  booktitle = {2018 {{IEEE}} European Symposium on Security and Privacy},
  year = 2018,
  pages = {288--306},
  doi = {10.1109/eurosp.2018.00028},
  author = {Wright, Charles and Varia, Mayank}
}

@article{Zalnieriute.2022bb,
  title = {{\emph{Big }}{{{\emph{Brother Watch}}}}{\emph{ and Others v. the }}{{{\emph{United Kingdom}}}}},
  year = 2022,
  journal = {American Journal of International Law},
  volume = {116},
  number = {3},
  pages = {585--592},
  doi = {10.1017/ajil.2022.35},
  author = {Zalnieriute, Monika}
}

@article{Zinkus.2022cc,
  title = {Cryptographic Confidentiality of Data on Mobile Devices},
  year = 2022,
  journal = {Proceedings on Privacy Enhancing Technologies},
  volume = {2022},
  pages = {586--607},
  doi = {10.2478/popets-2022-0029},
  
  author = {Zinkus, Maximilian and Jois, Tushar M. and Green, Matthew}
}

@misc{Zinkus.2022ds,
  title = {Data Security on Mobile Devices: Current State of the Art, Open Problems, and Proposed Solutions},
  year = 2021,
  number = {2105.12613},
  eprint = {2105.12613},
  publisher = {arXiv},
  doi = {10.48550/arXiv.2105.12613},
  archiveprefix = {arXiv},
  
  author = {Zinkus, Maximilian and Jois, Tushar M. and Green, Matthew}
}

@techreport{Zittrain:2016uv,
  title = {Don't Panic: Making Progress on the "Going Dark" Debate},
  year = 2016,
  number = {2016-1},
  institution = {Berkman Center for Internet \& Society at Harvard Law School},
  author = {Zittrain, Jonathan L. and Olsen, Matthew G. and O'Brien, David and Schneier, Bruce},
  note = {Retrieved from http://nrs.harvard.edu/urn-3:HUL.InstRepos:28552576},
  type = {Report}
}

@phdthesis{APLM.2022pc,
  type = {{PhD dissertation}},
  title = {{Direito \`a privacidade, criptografia e arquitetura democr\'atica} \emph{[The right to privacy, cryptography, and democratic architecture]}},
  year = 2022,
  address = {S\~ao Paulo},
  school = {University of S\~ao Paulo},
  author = {Monteiro, Artur Pericles Lima},
  collaborator = {{Virg\'ilio Afonso da Silva}}
}


\end{document}